\documentclass[final,1p,times]{elsarticle}
\biboptions{sort&compress}

\usepackage{amssymb}
\usepackage{amsmath}

\usepackage{tabularx}
\usepackage{lineno}
\usepackage{booktabs}
\usepackage{xurl}

\begin{document}

\begin{frontmatter}



\title{A Comprehensive Survey of Blockchain Scalability: Shaping Inner-Chain and Inter-Chain Perspectives}


\author[1]{Baochao Chen\fnref{fn}}
\ead{cbcchenbaochao@163.com}

\author[1]{Liyuan Ma\fnref{fn}}
\ead{mly@tju.edu.cn}

\author[1]{Hao Xu\fnref{fn}}
\ead{hao\_xu@tju.edu.cn}

\author[1]{Juncheng Ma\fnref{fn}}
\ead{juncheng@tju.edu.cn}

\author[1]{Dengcheng Hu\fnref{fn}}
\ead{hdc@tju.edu.cn}

\author[1]{Xiulong Liu\corref{cor}}
\ead{xiulong\_liu@tju.edu.cn}

\author[2]{Jie Wu}
\ead{jiewu@temple.edu}

\author[1]{Jianrong Wang}
\ead{wjr@tju.edu.cn}

\author[1]{Keqiu Li}
\ead{keqiu@tju.edu.cn}

\affiliation[1]{organization={Tianjin University},
            addressline={Haihe Education Park, Jinnan District},
            city={Tianjin},
            postcode={300354},
            country={China}}
\affiliation[2]{organization={Temple University},
            addressline={1801 N. Broad Street},
            city={Philadelphia},
            state={PA},
            postcode={19122},
            country={USA}}
\cortext[cor]{Corresponding author}
\fntext[fn]{The first five authors contribute equally to this survey.}

\begin{abstract}
Blockchain is widely applied in logistics, finance, and agriculture. As single blockchain users grow, scalability becomes crucial. However, existing works lack a comprehensive summary of blockchain scalability. They focus on single chains or cross-chain technologies. This survey summarizes scalability across the physical and logical layers, as well as inner-chain, inter-chain, and technology dimensions. The physical layer covers data and protocols, while the logical layer represents blockchain architecture. Each component is analyzed from inner-chain and inter-chain perspectives, considering technological factors. The aim is to enhance researchers' understanding of blockchain's architecture, data, and protocols to advance scalability research.
\end{abstract}

\begin{keyword}
Blockchains \sep Scalability \sep Inner-chain \sep Inter-chain
\end{keyword}

\end{frontmatter}


\section{Introduction}
\label{introduction}

\subsection{Background}
Blockchain technology has gained significant attention in recent years as a promising solution to various application issues, such as security, trust, and transparency. As a distributed ledger technology, it enables peer-to-peer transactions without intermediaries. However, scalability remains a critical challenge for blockchain technology in many real-world applications.

The scalability of a blockchain refers to its ability to maintain a certain level of performance and security while handling a growing volume of transactions. It depends on its architecture, consensus mechanism, block size, and transaction throughput. Therefore, it is necessary to explore the concept of blockchain scalability, evaluate different scalability methods and technologies, and analyze the performance of existing blockchain solutions in terms of scalability.

The scalability of a blockchain remains a significant concern, as blockchain networks are intended to process a substantial volume of transactions and accommodate an expanding user population. Applications such as financial services, supply-chain management, and decentralized platforms place increasing emphasis on the need for high transaction throughput and low latency. Consequently, addressing scalability has become imperative. In domains such as finance, efficient real-time handling of transactions is crucial to meet the requirements of swift payment and settlement  processes. Similarly, in supply-chain management, seamless processing of cross-organizational transactions is indispensable. Moreover, decentralized platforms, which are significant application areas of blockchain systems, impose even greater scalability demands to support extensive user interactions and facilitate smart contract execution.

\begin{figure*}[t]
  \centering
  \includegraphics[width=0.8\linewidth]{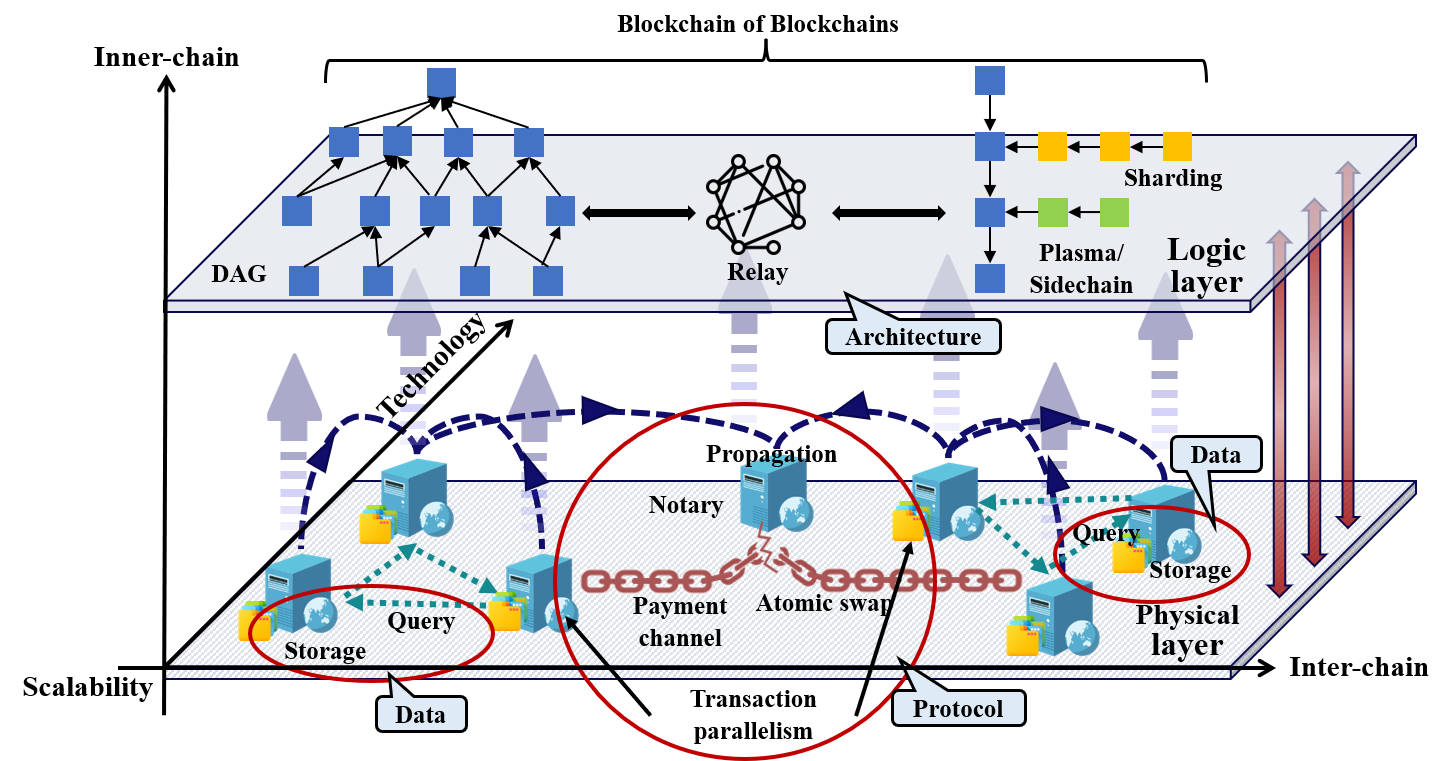}
  \caption{The overview of Blockchain.}
  \label{Blockchain}
\end{figure*}

Multiple studies have long focused on the scalability of blockchain, considering it as a key factor in improving performance. These studies primarily concentrate on the scalability within a single blockchain. But as the number of blockchains increases, cross-chain interoperability should also be considered as an aspect of scalability. Some research has summarized the technologies that enable scalability, including architecture, storage, and cross-chain capabilities. These studies largely emphasize technological advancements but do not deeply explore the importance of scalability at the system level. Therefore, a comprehensive approach is needed to address the scalability challenges faced by blockchain systems. Different from existing work, this survey summarizes the scalability of blockchain on two levels: the physical layer and the logical layer, and three dimensions: inner-chain, inter-chain, and technology. The physical layer consists of data and protocols, while the logical layer represents the architecture of the blockchain. Within each layer, we elaborate on each part from both inner-chain and inter-chain perspectives and incorporate the technology dimension into it. The detailed description is shown in Fig. \ref{Blockchain}.



\subsection{Related Work}
Blockchain system scalability has been a subject of scrutiny in the research community for a long time. 
Numerous surveys~\cite{kolb2020core, huang2021survey, ruan2021blockchains, cao2022blockchain, dotan2021survey} extensively analyzed blockchain systems, highlighting the importance of scalability and regarding it as an important research direction for enhancing the performance of blockchain systems.
\cite{nasir2022scalable} concentrates on scalability as its core theme, with an emphasis on innovative techniques and mechanisms aimed at improving the scalability of blockchain systems. Additionally, it evaluates numerous blockchain solutions in terms of scalability.
\cite{xie2019survey} similarly focuses on scalability and portrays scalability from three perspectives: throughput, storage, and networking.
These surveys primarily concentrate on describing scalability within a single blockchain system.
However, from our perspective, as the number of blockchain systems increases, facilitating interactions between different blockchains should also be considered as an aspect of scalability.

Achieving scalability within the blockchain requires the utilization of diverse technologies, and numerous studies have provided comprehensive summaries of these technologies.
Sharding and DAG have always been the core solutions for scalability from the architectural perspective.
Xi \emph{et al.}~\cite{xi2021comprehensive} focused on sharding as their core concept, emphasizing the latest research developments in sharded blockchain systems, encompassing shard configuration and cross-shard transaction processing. 
The study in~\cite{wang2023sok}, focused on DAG as a fundamental technology, analyzing the trade-offs between distinct factors, discussing open challenges, and examining the potential of using DAG-based solutions to advance scalability, and suggesting promising future research directions. 
As the amount of data stored in blockchain systems increases steadily, ensuring efficient storage and retrieval is a crucial component of blockchain scalability. 
Therefore, this article focused on addressing the prevention of excessive space consumption while maintaining rapid data retrieval. 
Some related studies have targeted these issues~\cite{wei2022survey, rottenstreich2021sketches}, emphasizing ways to reduce storage costs and improve query efficiency in blockchain systems. 
To clarify, prior studies on scalability had predominantly focused on single-chain scenarios. 
Nevertheless, as blockchain technology continues to see constant advancements, instances of data silos between chains are becoming increasingly prevalent. To address this issue, cross-chain technology aims to eliminate data barriers so as to facilitate interoperability between diverse blockchain systems~\cite{belchior2021survey, wang2023exploring, ren2023interoperability}. Therefore, cross-chain solutions may be viewed as a means to achieve inter-chain scalability.
These surveys offer a comprehensive account of specific technological advancements and serve as a valuable reference. However, their focus remains primarily on technical aspects, rather than delving into the significance of scalability at the system level.

In conclusion, we consider that the current related surveys offer only a limited overview of scalability, underscoring the necessity for a more comprehensive and inclusive approach toward addressing the scalability challenges faced by blockchain systems.



\subsection{Overview of Paper Structure}

\begin{figure*}[t]
  \centering
  \includegraphics[width=0.9\linewidth]{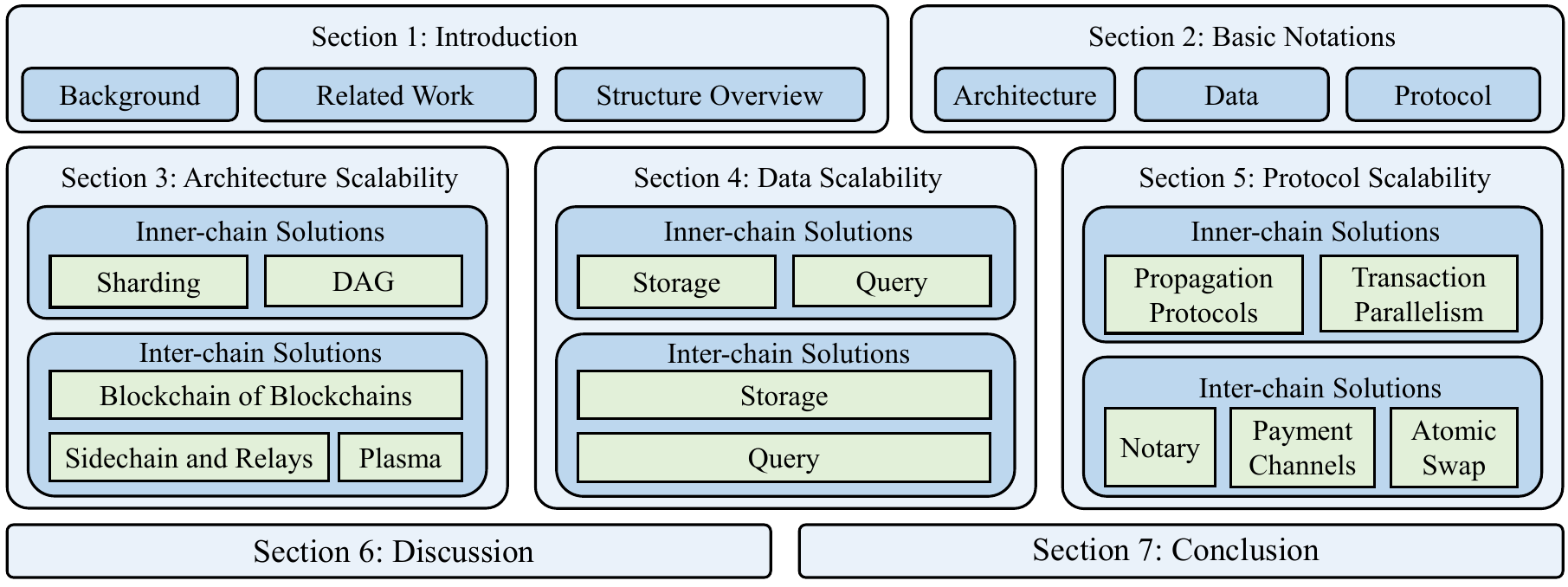}
  \caption{The overview of structure in this survey.}
  \label{structure_fig}
\end{figure*}

This paper provides a unique perspective to analyze the scalability of the existing blockchain approaches. 
We deconstruct the traditional six-layer blockchain architecture (data, network, consensus, incentive, contract, and application), from the perspective of scalability.
We propose that the three most important components that affect the scalability of a blockchain system are: the architecture at the logical layer and data and protocols at the physical layer.
Furthermore, we innovatively analyze the architecture, data, and protocols of blockchain systems from the inner-chain, inter-chain, and technology perspectives.

The contributions of this paper are mainly in the following three aspects:
\begin{itemize}
    \item We innovatively analyze and sort out the existing state-of-the-art work from the three perspectives of architecture at the logical layer and data and protocols at the physical layer.
    \item We classify the existing work from the inner-chain, inter-chain, and technology perspectives to systematically analyze the approaches that improve blockchain scalability.
    \item Based on a comprehensive summary of the work on blockchain scalability, and through open discussions, this survey gives a landscape of the future development of blockchain scalability.
\end{itemize}




The organization of this survey is shown in Fig. \ref{structure_fig}. Section \ref{basic} introduces the basic notations used in this paper. Section \ref{architecture} analyzes the technologies that improve scalability in terms of architecture-mainly, inner-chain technologies such as sharding and DAG, and inter-chain technologies such as blockchain of blockchains, sidechain and relays, and plasma. Section \ref{data} introduces technologies to improve data scalability, including inner- and inter-chain storage and querying. Section \ref{protocol} presents the technologies related to protocol scalability improvement, including inner-chain propagation protocols, transaction parallelism, inter-chain notaries, payment channels, and atomic swaps. 
Section \ref{discussion} gives an open discussion of blockchain scalability and gives a landscape of potential blockchain scalability improvements.
Section \ref{conclusion} concludes this survey.


\section{Basic Notations}
\label{basic}
In recent years, blockchains have been applied in many fields such as logistics, finance, and agriculture. With the increase in the number of users, the scalability of blockchain has become an inevitable issue and may become a bottleneck to further development. Therefore, in this survey, we focus on the existing works on improving the scalability of blockchains. Based on the summary and analysis of the literature, we innovatively divide the existing works based on three aspects from the inner-chain and inter-chain perspectives: architecture scalability, data scalability, and protocol scalability. Before delving into the details of each aspect, to help understand this survey better, we present some notations that we use in the following sections.

\subsection{Inner-chain and Inter-chain}
Scalability is a pivotal factor in shaping blockchain technology's success and widespread adoption. Innovatively, we analyze its scalability from both inner-chain and inter-chain perspectives, which encompass the majority of research efforts in existing blockchain systems. Inner-chain scalability primarily pertains to enhancing the performance and capacity of a single blockchain. This can be achieved through the adoption of architectures like sharding, and DAG, the design of efficient storage and query strategies, and the development of propagation protocols and transaction parallelism strategies. Inter-chain scalability, on the other hand, refers to the ability to coordinate and process transactions across different blockchains. Technologies such as sidechain/relay chain, off-chain access, and atomic swap facilitate interoperability among diverse blockchains, facilitating asset and data exchange and achieving broader scalability. By comprehensively considering these two aspects, the blockchain ecosystem can adapt to the growing demands, providing users with an efficient, secure, and scalable distributed ledger technology, thereby establishing a robust foundation for future decentralized applications.

\subsection{Architecture Scalability}
As the number of users and their transaction volumes increase, it is necessary to speed up the processing of transactions in blockchains and achieve an overall increase in the network throughput. Therefore, we define the system's ability to grow and adapt to the increasing transaction demand as the architecture scalability. There are two aspects to architecture scalability: inner-chain scalability and inter-chain scalability. Inner-chain scalability of architecture refers to the means of parallelly processing user transactions through a distributed network of nodes within the blockchain, and mainly includes sharding and DAG techniques. The fundamental concept behind sharding is to partition the network into multiple groups, referred to as committees, that work concurrently to handle transactions. DAG breaks the chain structure of the blockchain and constructs transaction blocks into a graph topological structure, allowing the verification of multiple transactions simultaneously. Compared to single-chain expansion, we define the inter-chain scalability of the architecture as a means of introducing a homomorphic/heterogeneous blockchain based on the existing blockchain structure. Its purpose is to accelerate the transaction processing and realize more extensive and complex application-scenario requirements. There are several methods for achieving inter-chain scalability, such as blockchain of blockchains, sidechain and relay, and plasma. Blockchain of blockchains is a cross-chain technology that can connect different blockchain networks to form a larger and stronger blockchain network. A sidechain is a blockchain parallel to the main chain, which can perform functions and applications different from that of the main chain. A relay is a mechanism that bridges different blockchains, enabling transaction interoperability between different blockchains. Plasma is an Ethereum on-chain scaling technology. The core idea is to perform complex calculations on a sidechain to improve the performance and throughput of the Ethereum main chain. This achieves higher processing capacity and lower costs.

\subsection{Data Scalability}
In a blockchain network, nodes must maintain a complete ledger of data, and an increase in the transaction volume will inevitably increase the cost of data storage for the nodes. Additionally, when users need to retrieve a certain part of the data, the cost of querying will also increase. Many studies have attempted to increase the throughput of the blockchain network. However, improving the throughput will increase the burden on the blockchain nodes, in terms of storing and querying massive amounts of data. This hinders some resource-limited devices (such as mobile phones and tablets) from joining the blockchain network. In extreme cases, the number of nodes that can bear the burden of storing and querying data in the network gradually decreases, causing the blockchain network to evolve from a distributed network to a centralized network. This clearly violates the decentralized nature of blockchains. Therefore, we define the concept of data scalability, which refers to the ability of the nodes in a blockchain network to reduce data storage and querying costs while satisfying the availability requirements of the blockchain. We analyze the data scalability from both inner-chain (on-chain) and inter-chain (off-chain and multi-chain) perspectives. Here, on-chain nodes optimize key information such as data storage, data indexing, and related hash values. This helps to reduce the node resource consumption and improve the storage and querying capabilities of the overall system. We refer to this process as inner-chain scalability. Inter-chain scalability involves storing detailed data and its data structure in off-chain nodes. Inner-chain indexing and hash values are then used to achieve fast location and verification of inter-chain data. This process improves the transaction execution speed and overall network throughput, addressing the scalability challenges of blockchain data storage and querying. In addition, we extend the inter-chain node to the concept of multi-chain nodes to more comprehensively cover data scalability. For a better description, the on-chain is collectively referred to as the inner-chain, and the off-chain and multi-chain interactions are called inter-chain.

\subsection{Protocol Scalability}
The operation of a blockchain system relies on multiple protocols, each of which significantly impacts the system's reliability and performance. 
To maintain the efficiency and reliability of transaction processing in the blockchain system when expanding, it is necessary to extend the various protocols supporting its operation. 
Thus, we define protocol scalability as the ability of a blockchain system to meet large-scale and high-concurrency requirements without compromising transaction execution speed and system reliability. 
This scalability encompasses inner-chain protocol scalability and inter-chain protocol scalability. 
Specifically, inner-chain protocol scalability pertains to optimizing protocols such as broadcasting and transaction parallelism in the blockchain system to enhance reliability and performance. 
The efficiency of network data transmission is crucial for performance, with the broadcasting protocol facilitating information transmission between nodes. 
An efficient broadcasting protocol can prevent malicious attacks, improve the reliability and performance of the blockchain system, and enhance system scalability. 
Transaction execution involves each node adding new transaction data to its local ledger according to specific rules. 
Multiple transactions are processed and verified simultaneously by nodes, known as transaction parallelism. By processing transactions in parallel, the speed of transaction confirmation can be accelerated, the processing time can be reduced, and the overall performance of the blockchain system can be enhanced. 
Protocol inter-chain scalability refers to the interoperability and stability of interaction between blockchains, and includes techniques such as notary systems, payment channels, and atomic swaps. 
Notary systems introduce multiple notaries to act as trusted intermediaries for cross-chain transactions, ensuring transaction reliability and tamper resistance. 
Payment channels are peer-to-peer transaction models based on blockchain technology, which achieve fast and low-cost transaction processing by establishing direct connection channels between participants. 
Atomic swaps are peer-to-peer cross-chain transaction methods that guarantee the security, reliability, and irreversibility of transactions between two chains.

\section{Architecture Scalability}
\label{architecture}

The current business environment presents a challenge for the blockchain system to be widely embraced as a feasible alternative to traditional databases, primarily because its transaction throughput capacity falls short of meeting the essential transaction processing demands.
To tackle this limitation, various\textit{ Inner-chain}  and \textit{Inter-chain} architectural scaling techniques have gained prominence.
In this Section, we showcase the architecture scalability of blockchain from Inner-chain and Inter-chain perspectives. As shown in  Table \ref{shard_tab and dag_tab},  Inner-chain concludes sharding and DAG. At this stage, technologies of sharding and DAG are mainly used to improve the scalability of the architecture and this survey provides a detailed study. 
We categorize technologies of Inter-chain into three types: sidechain \& relay, plasma, and blockchain of blockchains (BoBs).
The following subsections will demonstrate them one by one.


\subsection{Inner-chain Solutions}

\begin{figure*}[t]
  \centering
  \includegraphics[width=0.9\linewidth]{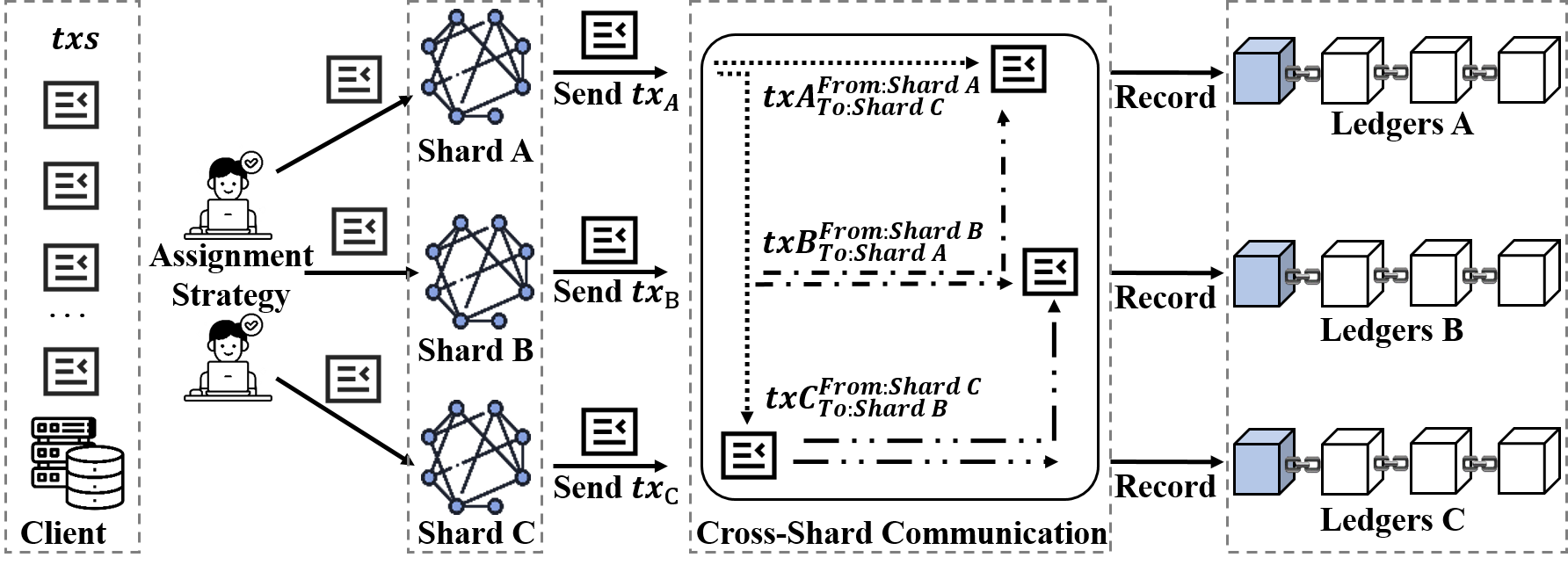}
  \caption{The overview of sharding scalability.}
  \label{sharding_fig}
\end{figure*}

\subsubsection{Sharding}
The scalability limitations of blockchain technology hinder its application in various scenarios, such as e-commerce and freight transport. Low throughput is a pressing challenge that needs to be addressed. Sharding is one of the effective solutions for scalability issues in blockchain technology. Therefore, both academic and industrial communities have made significant efforts toward sharding scalability.  As shown in Fig. \ref{sharding_fig}, the central tenet of sharding entails the segmentation of the network into distinct committees that operate independently, enabling simultaneous processing of transactions. This mechanism enhances the system's overall throughput and efficiency.
Many novel sharding protocols have emerged in recent years.

The sharding work has some novel architecture designs, such as network sharding, multi-layer design, double-chain architecture, etc., providing a source of innovation for the continuous innovation of sharding work. Luu \emph{et al.}~\cite{luu2016secure} studied a new distributed agreement protocol in permissionless blockchains. Their method scaled the transaction rates almost linearly with respect to the capability of miners by uniformly partitioning the mining network into shards. Omniledger~\cite{kokoris2018omniledger} designed a scale-out distributed ledger that maintained long-term security within a permissionless operation. This method introduced atomic commit protocol (Atomix) to commit transactions atomically across shards. RapidChain~\cite{zamani2018rapidchain} presented a novel public blockchain protocol designed to address the scalability and security constraints. This protocol incorporates an efficient consensus algorithm within each committee, ensuring optimal throughput by leveraging block pipelining techniques. Monoxide~\cite{wang2019monoxide} provided a scalable blockchain system that allows for the independent representation and parallel execution of workloads in communication, computation, and storage. It implemented Chu-ko-nu mining as a countermeasure to maintain the security threshold even in situations where mining power is dispersed across multiple zones. Pyramid~\cite{hong2021pyramid} is a novel layered sharding system that departs from the traditional approach of isolating shards completely. Instead, it allows shards to overlap, facilitating collaboration and coordination among them. To achieve this, a layered sharding consensus protocol is introduced, enabling the seamless commitment of cross-shard blocks within each shard. This protocol leverages collective efforts and cooperation across different shards, resulting in improved scalability and enhanced system performance.
RepChain~\cite{huang2020repchain}, is a reputation-based blockchain system that utilizes sharding to enhance security and speed. It introduces a unique double-chain architecture, combining CSBFT consensus for reputation tracking and Raft consensus for high throughput. This approach encourages node collaboration and improves scalability, making it a significant advancement in sharding-based systems.

Many studies have proposed various sharding strategies, including sharding reconfiguration, dynamic node joining and leaving, and selection of the best shard node. OptChain~\cite{nguyen2019optchain} optimizes transaction placement within shards to enhance the performance of existing sharding approaches. It is designed to be flexible and compatible with various sharding methods, allowing for efficient allocation of transactions and ultimately improving the overall system performance.
Huang \emph{et al.}~\cite{huang2022elastic} studied the allocation of budget-limited network resources to shards in a practical Byzantine fault tolerance (PBFT) based permissioned blockchain. 
Their work proposed a novel algorithm based on the drift-plus-penalty approach, aiming to achieve a resource-allocation solution that is close to optimal. Crain \emph{et al.}~\cite{crain2021red} proposed the Red Belly  Blockchain (RBBC), the first secure blockchain that could be scaled to hundreds of geo-distributed consensus nodes. This system offers a new balancing method that totally orders all transactions while assigning them to distinct roles. PolyShard~\cite{li2020polyshard}, is an innovative approach that addresses the challenges of achieving linear scalability in terms of throughput, storage efficiency, and security. Drawing inspiration from coded computing, specifically Lagrange Coded Computing, PolyShard introduces a novel concept. Rather than storing and processing a single uncoded shard, as traditionally done, each node in PolyShard operates on a coded shard of equivalent size, enabling enhanced storage and computational capabilities.
Metosis~\cite{marson2021mitosis} addresses a new class of problems, namely, when and how nodes dynamically join and leave the system, to achieve optimal system performance. Metosis implements dynamic effects of chains by creating, adding, splitting, and merging chains, and has been experimentally deployed in Fabric Chaincode. Gearbox~\cite{david2022gearbox} addresses the monolithic outage issue by dynamically adjusting the number of commission nodes. Specifically, the Gearbox consensus protocol runs a minimum number of nodes per shard, and the chain of control periodically receives "heartbeat transactions" sent by each shard to monitor shard activity. Once the shard becomes inactive, Gearbox immediately increases the number of nodes in the shard until it is activated again. Zhang \emph{et al.} proposed a deterministic and fast transaction allocation scheme TxAllo~\cite{zhang2022txallo}. By transforming a transaction assignment problem into a community detection problem in graph-structured data, TxAllo dynamically deduces the association between an account assignment and its transactions. The TxAllo protocol consists of global TxAllo and dynamic TxAllo.

\begin{table*}
    \caption{Inner-Chain Solutions of Architecture Scalability}
    \tiny
    \label{shard_tab and dag_tab}
    \begin{tabularx}{\textwidth}{ccX}
    \toprule
    Inner-Chain Solutions & Types & Solutions \\
    \midrule 
    Sharding & Account-Based Transactions & Pyramid\cite{hong2021pyramid} RepChain\cite{huang2020repchain} OptChain\cite{nguyen2019optchain} RBBC\cite{crain2021red} Metosis\cite{marson2021mitosis} Gearbox\cite{david2022gearbox} TxAllo\cite{zhang2022txallo} BrokerChain\cite{huang2022brokerchain} SharPer\cite{amiri2021sharper} LBChain\cite{li2023lb} Benzene\cite{cai2022benzene} ChainSpace\cite{al2017chainspace} Jenga\cite{li2022jenga} Meepo\cite{zheng2022meepo} Harmony\cite{harmony} Elrond\cite{elrond} Zilliqa\cite{zilliqa} Near\cite{near} Eth2.0\cite{eth2.0}     \\
    Sharding & Support for Smart Contract &  Chainspace\cite{al2017chainspace} Jenga\cite{li2022jenga} Meepo\cite{zheng2022meepo} Harmony\cite{harmony} Elrond\cite{elrond} Zilliqa\cite{zilliqa} Near\cite{near} Eth2.0\cite{eth2.0}  \\
    Sharding & Industry Sharding System &  Harmony\cite{harmony} Elrond\cite{elrond} Zilliqa\cite{zilliqa} Near\cite{near} Eth2.0\cite{eth2.0}  \\       
    DAG & Divergence & IOTA\cite{anakath2022tangles} Graphchain\cite{boyen2018graphchain} Meshcash\cite{bentov2021tortoise} Spectre\cite{sompolinsky2016spectre} Avalanche\cite{rocket2018snowflake}  \\
    DAG & Parallel & Nano\cite{lemahieu2018nano} Dlattice\cite{zhang2019dledger} Jointgraph\cite{xiang2021jointgraph} Chainweb\cite{chainmaker} Aleph\cite{gkagol2018aleph} Vite\cite{liu2018vite} Dexon\cite{chen2018dexon} Hashgraph\cite{baird2016swirlds}\\
    DAG & Convergence & Byteball\cite{churyumov2016byteball} Conflux\cite{li2018scaling} Tips\cite{chen2022tips} Nezha\cite{xiao2022nezha}\\    
    \bottomrule
    \end{tabularx}
\end{table*}

Cross-shard transactions are a natural consequence of sharding systems, and the uneven distribution of cross-shard transactions can undermine the performance of the sharding system. Brokerchain~\cite{huang2022brokerchain} is a cross-shard protocol for account/balance-based state sharding. It aims to generate fewer cross-shard transactions and ensure workload balance for all shards. Sharper~\cite{amiri2021sharper} facilitates concurrent transaction processing through node clustering and the sharding of both data and the ledger, enabling parallelized operations. LBChain~\cite{li2023lb} proposes a new method to alleviate the problem of load imbalance by periodically migrating active accounts from heavily loaded shards to lightly loaded ones, achieving a dynamic balance of transactions between shards. LBChain uses LSTM for transaction prediction and account allocation. Based on the prediction results, the node network migrates the accounts along with their transactions, as a whole, to the new shard.

A trusted execution environment (TEE) can provide trusted security guarantees for a system by securely isolating sensitive computations and performing secure authentication of the execution environment. TEE has been adopted by some sharding systems to provide security services. The proposed scheme~\cite{dang2019towards} focuses on enhancing BFT consensus protocols by utilizing Trusted Execution Environments (TEE) to effectively eliminate equivocation in scenarios involving Byzantine failures. Benzene~\cite{cai2022benzene} effectively addresses the risk of a single-shard outage because of decentralized computing and reduces the node storage and validation overhead by using a collaborative consensus protocol in conjunction with a secure TEE. The overall architecture adopts a double-chain structure. The proposal chain is responsible for recording transactions, and the voting chain is responsible for the collaborative consensus of fragments.

In addition to classic transfer transactions, smart contract transactions, with their complex code logic, are capable of handling more complex scenarios, such as smart metering, complex voting, and privacy-protected banking transactions. A large number of systems can now support smart contract transactions. Chainspace~\cite{al2017chainspace} introduces a scalable system that exhibits unlimited scalability as the number of nodes grows. George \emph{et al.}~\cite{pirlea2021practical} introduced Cosplit, an advanced static analysis tool designed to accurately extract ownership and commutativity information from smart contract source code. This valuable information is then utilized to generate sharding signatures for enhanced performance and efficiency. Jenga~\cite{li2022jenga} proposed a system that orchestrates the state storage, logic storage, and execution of smart contracts, instead of treating the contract as an indivisible entity. Meepo~\cite{zheng2022meepo} introduces a novel methodology that encompasses a partial cross-call merging strategy, enabling smart contracts to facilitate flexible and concurrent invocations across multiple shards. This approach effectively addresses the requirements posed by intricate business models within a consortium-based blockchain system.

The industry has also seen the emergence of high-performance sharding systems, which are primarily designed around security and atomicity considerations. Harmony~\cite{harmony} is a next-generation sharding-based blockchain, which is fully scalable, provably secure, and energy efficient. Elrond~\cite{elrond} presents a  novel architecture that introduces a genuine state-sharding scheme for practical scalability, eliminating energy and computational waste while ensuring distributed fairness through an SPoS consensus. With scalability as the main goal, Zilliqa~\cite{zilliqa} proposes a new smart contract language, Scilla, which scales much better for a multitude of applications that range from automated auctions and shared economy to financial modeling. NEAR~\cite{near} is a decentralized application platform aiming at creating future open networks and empowering their economies. It employs the core foundational technology of Bitcoin and combines it with cutting-edge advancements in community consensus, database sharding, and availability. The main feature of  Eth2.0~\cite{eth2.0}  is the transition from PoW to PoS, which improves upon PoW by being much more scalable and accessible. Sharding will help in scaling up Eth2.0 throughput exponentially, by breaking down data verification into smaller shards and enabling parallel processing.

\subsubsection{DAG}
The high latency and low scalability of traditional blockchain systems limit their wide application in a variety of scenarios. DAG is an effective technique that can overcome this limitation. The core principle of DAG is to construct transactions in the form of graph topologies, replacing the traditional linear block structure, which allows multiple transactions to be validated simultaneously, thus improving the transaction processing speed and overall network throughput. In recent years, a variety of DAG blockchains have emerged; however, there is still a lack of a systematic summary of the works on DAG implementation scalability.

Based on the DAG formed by the graph topology, it can be summarized into three types: divergence, parallel, and convergence. 
To elaborate, divergence refers to a network expanding in an 
unpredictable direction, where the order cannot be predetermined.
In a parallel topology, multiple chains are maintained concurrently. 
Convergence implies that the blockchain network tends to be organized in a 
sequential order.

We first explain divergence. 
Divergence DAGs provide higher parallelism as blocks can be added in any order. 
This makes divergent DAGs well-suited for high-throughput applications such as large-scale transaction processing or distributed storage systems. 
IOTA~\cite{anakath2022tangles}  is a permissionless network where each node can freely participate or leave. 
%
The main innovation is a distributed ledger structure based on DAG, called Tangle, which is a blockchain with no blocks or chains.
%
%
Graphchain~\cite{boyen2018graphchain} is formed by executing each transaction to confirm its ancestry. 
%
%
In contrast, Graphchain is different from IOTA in terms of the incentive mechanism. 
IOTA operates under tail-selection rules and is not affected by excitation. 
Instead, Graphchain introduces an incentive mechanism for maintaining graphs. 
Meshcash~\cite{bentov2021tortoise} is a hierarchical DAG system where an honest node generates new blocks via PoW, referencing all the end blocks in its view. Each block contains a level number. 
Meshcash provides a simple, but limited security approach that is complex and attack-resistant, using an inter-chain asynchronous Byzantine protocol. 
Spectre~\cite{sompolinsky2016spectre} uses a DAG structure for faster block generation and larger block capacity. 
The performance improvement mainly comes from two aspects: first, the system structures the blocks to form a topological network. 
Transactions can be added to the network simultaneously, which makes the system scalable. Second, increasing the block generation rate helps improve the performance because Spectre requires only paired sorting between two blocks, avoiding the obstacle of many conflicting states between blocks. 
Avalanche~\cite{rocket2018snowflake} is a public chain system with a new consensus. 
Unlike the BFT class and Nakamoto mechanism, Avalanche uses Slush, a CFT fault-tolerant mechanism, as its underlying protocol. 
%
%
Finally, an enhancement algorithm is applied to the whole topological DAG network.

The second class covers parallel DAGs. 
Parallel DAGs provide high security as multiple chains remain parallel.
This means that even if one chain is attacked, other chains can maintain integrity. 
Additionally, parallel DAGs are suitable for processing transactions or data with the same priority. 
%
Nano innovatively~\cite{lemahieu2018nano} adopts the method of one user, one chain. It records only its own transactions, and only it can modify its records. It does not share data with other accounts; thus, all transactions can be executed in parallel, providing transaction speeds in the range of seconds and infinite scalability. 
%
Hashgraph~\cite{baird2016swirlds} is a permissioned network. 
Hashgraph has been pioneering for asynchronous BFT consensus in the public chain environment. A major problem of traditional BFT is the high complexity of messages, which results in high network bandwidth consumption and failure to cope well with the dynamic network. 
%
DLattice~\cite{zhang2019dledger} uses the so-called DPOS-BA-DAG protocol to reach consensus. DPOS provides a way for committees to form, and BA shows how consensus can be achieved in DAGs. 
Jointgraph~\cite{xiang2021jointgraph} simplifies the voting process to one round by introducing supervisory nodes. These nodes replace the misbehaving node with an honest node, monitor the nodes, and periodically take snapshots of the system status to release memory. Specifically, each transaction in Jointgraph is broadcast to its peers through the gossip protocol. 
%
Chainweb~\cite{martino2018chainweb} is a permissionless system that attempts to scale the Nakamoto consensus by maintaining multiple parallel chains. 
%
Aleph~\cite{gkagol2018aleph} enables each node to publish messages equally and concurrently, and is represented as a cell designed to transmit asynchronously and efficiently across the network. 
Each cell is independent and free to create, propagate, and vote. The core is to establish an overall ranking among these cells. 
Vite~\cite{liu2018vite} follows the basic structure of Nano, but introduces a global snapshot chain, a consistent storage structure, to achieve a total sequential sequence. Each account in Vite creates separate parallel transactions. 
The purpose of the snapshot block is to store the state of the Vite ledger. 
%
Dexon~\cite{chen2018dexon} has several parallel blockchains, each of which agrees independently. 
The consensus part is mainly divided into two modules. One is the single-chain consensus protocol.  The other is the butyl parallel chain to sort blocks.

The last class involves convergence DAGs. 
Convergence DAGs provide ordered transaction processing as blocks are added in a certain order. 
This makes convergent DAGs well-suited for applications that require temporal ordering, such as log recording and timestamping. 
Furthermore, owing to the ordering of blocks, convergent DAGs can also provide better data compression and storage efficiency. 
Byteball~\cite{churyumov2016byteball} is a permissionless network. The concept of a main chain and witness is innovatively introduced to encourage the verification of multiple parent transaction units, forming a digital signature Hash network with the growth of transactions, mutual verification, and enhanced security. 
%
Conflux~\cite{li2018scaling} inherits the Ghost design to achieve high performance without compromising on security. The main contribution is to decouple block confirmation from transactions. 
%
TIPS~\cite{chen2022tips} proposes a transaction inclusion detection protocol with a tagging signal. By generating transaction and block association information through a Bloom filter located in the block header, the block header is broadcast with a tagging signal priority, and other miners adjust their transaction packaging strategy accordingly upon receiving the signal. Through this signaling mechanism, the number of transaction conflicts in a block can be reduced. Furthermore, this approach can resist denial-of-service attacks and delay attacks. 
NEZHA~\cite{xiao2022nezha} proposes an efficient concurrency control scheme for blockchains based on DAG, which resolves the problems arising from conflicting concurrent read and write operations to the same address during the parallel processing of transactions. Specifically, first, a conflict graph is built based on the transaction addresses, to generate the overall order of transactions. Then, a hierarchical sorting algorithm is designed, which deduces the sorting level of each address from the conflict graph and sorts the transactions at each address.

\subsection{Inter-chain Solutions}
%
%
%
%
%
%

\subsubsection{Sidechain and Relay}
Sidechains are autonomous blockchains that are not standalone platforms; instead, they are linked to the main chain in a specific manner. 
Interoperability is a key feature of sidechains, allowing assets to move freely between the main chain and the sidechain.
Various methods can be employed to ensure seamless fund transfers. 
For instance, it is possible to deposit funds into a designated address and subsequently shift assets from the main chain to the side chain. During this process, the funds remain locked at the address, while the side chain reflects the corresponding amount. 
Alternatively, a more direct approach involves sending funds to a custodian who then carries out the exchange, converting the assets into the corresponding margin on the sidechain.

The work by Gaži P \emph{et al.}~\cite{gavzi2019proof} marked a significant milestone by introducing the first formal definition of a sidechain system. 
Their research elucidated the safe transfer of assets between sidechains and introduced a novel security definition.
This security definition extended the conventional transaction ledger properties of liveness and safety to encompass multiple ledgers.
Importantly, it introduced a "firewall" security property, fortifying each blockchain against potential sidechain failures.
Singh A \emph{et al.}~\cite{singh2020sidechain} made notable contributions by conducting an exhaustive analysis of sidechains and platforms.
Their comprehensive review encompassed recent developments in the field and offered a multi-faceted perspective on their impact.
Additionally, they critically highlighted the limitations of existing solutions and proposed innovative approaches to enhance the overall blockchain system.
Kiayias A \emph{et al.}~\cite{kiayias2020proof} advanced the field by creating the first sidechain architecture that enables direct communication between Proof of Work (PoW) blockchains.
They introduced the concept of a "two-way peg" to facilitate the transfer of assets between different chains.
Their work emphasized the prerequisites for inter-chain communication, which include a PoW blockchain as the source and a smart contract-capable blockchain as the destination. 
Moreover, they provided detailed insights into the required smart contracts for the implementation of these interlinked sidechains.

The relay serves as an intermediary or bridge between different blockchains or blockchain layers within a multi-chain ecosystem.
BTC-Relay~\cite{relay2020bridge} stands as a groundbreaking achievement, serving as the inaugural bridge between the Bitcoin blockchain and Ethereum smart contracts. 
Frauenthaler P \emph{et al.}~\cite{frauenthaler2020eth} have significantly reduced the operational costs associated with Ethereum-based blockchain relays, with potential cost reductions of up to 92\%.
Their pioneering approach combines a validation-on-demand pattern with an incentive structure, making decentralized interoperability between blockchains, such as Ethereum and Ethereum Classic, a practical reality.
In order to create and communicate with various types of sidechains without being aware of their fundamental structure, Garoffolo A \emph{et al.}~\cite{garoffolo2020zendoo} provided a construction technique for blockchain systems, similar to that of Bitcoin. 
They have implemented a universally verifiable transfer mechanism for sidechains, leveraging zk-SNARKs and sidechain nodes. 
Importantly, this mechanism allows sidechain nodes to directly witness the mainchain, while mainchain nodes only need to verify cryptographically validated certificates provided by sidechain maintainers.
This innovative approach enhances security and trust in the cross-chain ecosystem.

\subsubsection{Plasma}

The plasma~\cite{poon2017plasma} establishes a network of plasma subchains linked to the root chain, with the Merkel root of all transactions in all blocks of each subchain published to the root chain as a tool for verifying the data on the sidechain subsequently. 
This minimizes trust while allowing verifiable proof of fraud and an enforceable state.
By performing the aforementioned operation, a significant portion of the root chain's transaction load is transferred to the side chain for processing, requiring the root chain to only carry out verifiable forgery of transactions in the sidechain. 
This greatly improves the performance of the blockchain while processing transactions. 
M. H. Ziegler \emph{et al.}~\cite{ziegler2019integration} proposed a brand-new system architecture that uses the plasma framework to combine fog computing and blockchain technology and assess the effectiveness of its prototype.

\subsubsection{Blockchain of Blockchains}
\begin{figure*}[t]
  \centering
  \includegraphics[width=0.88\linewidth]{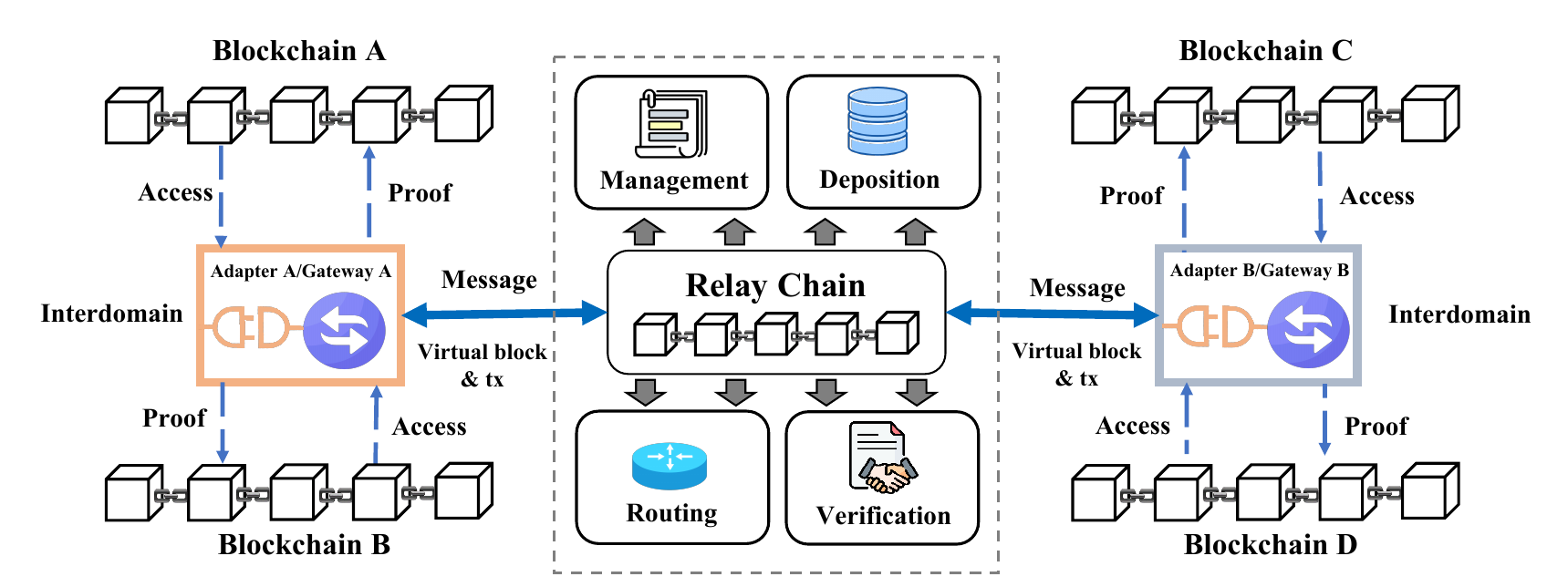}
  \caption{The overview of Blockchain of Blockchains.}
  \label{bobs_fig}
\end{figure*}

Different from the design of sidechains and subchains, BoBs attempt to restructure the existing blockchain inter-chain architecture to create a cross-chain Internet, as shown in Fig.~\ref{bobs_fig}.
Based on the types of blockchains within the ecosystem, BoBs can be categorized as public BoBs and consortium BoBs.

As a forerunner of public BoBs, Polkadot~\cite{wood2016polkadot} standardizes the method for passing messages between parallel chains of homogenous relay networks. 
It acknowledges the interoperability of parallel chains, which may engage in cross-chain interactions to reinforce the performance of one another. 
A standard created by Polkadot 2.0 called XCM~\cite{burdges2020overview} enables protocol designers to specify the types of data and sources from which their chains can send and receive data. 
It comes with one virtual machine that enables flexible execution and one virtual machine allows for adaptable execution. 
Another pioneer in the BoBs ecosystem is Cosmos. 
According to the IBC~\cite{kwon2019cosmos}, Cosmos is an end-to-end, connection-oriented, stateful protocol used to provide authenticated, reliable communication across diverse blockchains arranged in a dynamic topology. 
These advancements in the Polkadot ecosystem enhance the versatility and efficiency of cross-chain communication and execution, ultimately contributing to the broader development of blockchain technology.
Chainlink created CCIP~\cite{diaz2021oracle}, focusing on end-to-end security, futuristic interoperability, and an easy development process. 
The CCIP infrastructure consists of three layers: a message layer (programmable pass bridge), a transport layer (CCIP core), and a decentralized prophet network (DON) based on the OCR2.0 protocol. 
External developers only need to develop sender and receiver contracts, all other components are included in the CCIP service, and the developers can easily interact across chains through a unified interface.
LayerZero~\cite{zarick2021layerzero} is a trustless interoperability protocol, which provides a powerful, low-level communication primitive upon which a diverse set of cross-chain applications can be built.
Aion~\cite{spoke2017aion} enables the decentralized internet and supports a public cross-chain with the Aion Virtual Machine. 
Komodo~\cite{lee2018komodo} designs a three-in-one product that combines a wallet, cross-chain bridge, and decentralized exchange designed to be accessed through any Internet browser and connects to more than 60 blockchains, including Ethereum, Polygon, Avalanche, BNB Chain, and Cosmos.

As for consortium blockchains, ChainMaker~\cite{chainmaker} uses several components to complete a cross-chain operation, including a cross-chain agent, SPV, and transaction contract. 
The business contract needs to provide interfaces for forward and reverse operations. 
The forward interface is the contract portal to be invoked when the business is completed; the forward interface operation fails for business rollback.
XuperCross~\cite{wei2020xuperchain} solves the interoperability problem between heterogeneous blockchain systems (including public, private, and consortium chains) through the XIP protocol, which describes the cross-chain problem in an abstract manner and designs a standard solution. 
XIP contains a series of sub-protocols, including Naming Protocol, Cross Chain Transaction Consistency Protocol, and Data Authentication and Communication Protocol.
BitXHub~\cite{bitxhub} is the first cross-chain platform to support the W3C standard DID protocol, which is composed of three parts: relay chain, cross-chain gateway, and application chain. 
A common inter-chain transfer protocol, InterBlockchain Transfer Protocol, has been designed to allow heterogeneous assets, data, and services to be called across chains.
The BSN Interchain Communications Hub~\cite{ganesh2020identification} adopts a double-layer structure, utilizing relay chains as cross-chain coordinators, multiple heterogeneous chains as cross-chain transaction executors, and a relayer as a carrier of cross-chain data. 
Each application chain can verify the legitimacy of the cross-chain transactions on its own, thus ensuring the security of cross-chain interactions.
The source-oriented interoperability protocol known as Luyu~\cite{luyu} is a collection of adaptable, dependable, and unified interoperability protocols that enable the simple access and dependable operation of various reliable sources. 
Developers only need protocol-oriented programming to realize secure interaction with different trustworthy sources. 
Wecross~\cite{wecross} proposed four core technologies: UBI universal block link port, HIP heterogeneous chain interconnection protocol, TTM trusted transaction mechanism, and MIG multi-lateral cross-domain governance, which realizes efficient availability, security, and trustworthiness, and convenient governance of cross-chain interactions.

\begin{figure*}[t]
  \centering
  \includegraphics[width=0.9\linewidth]{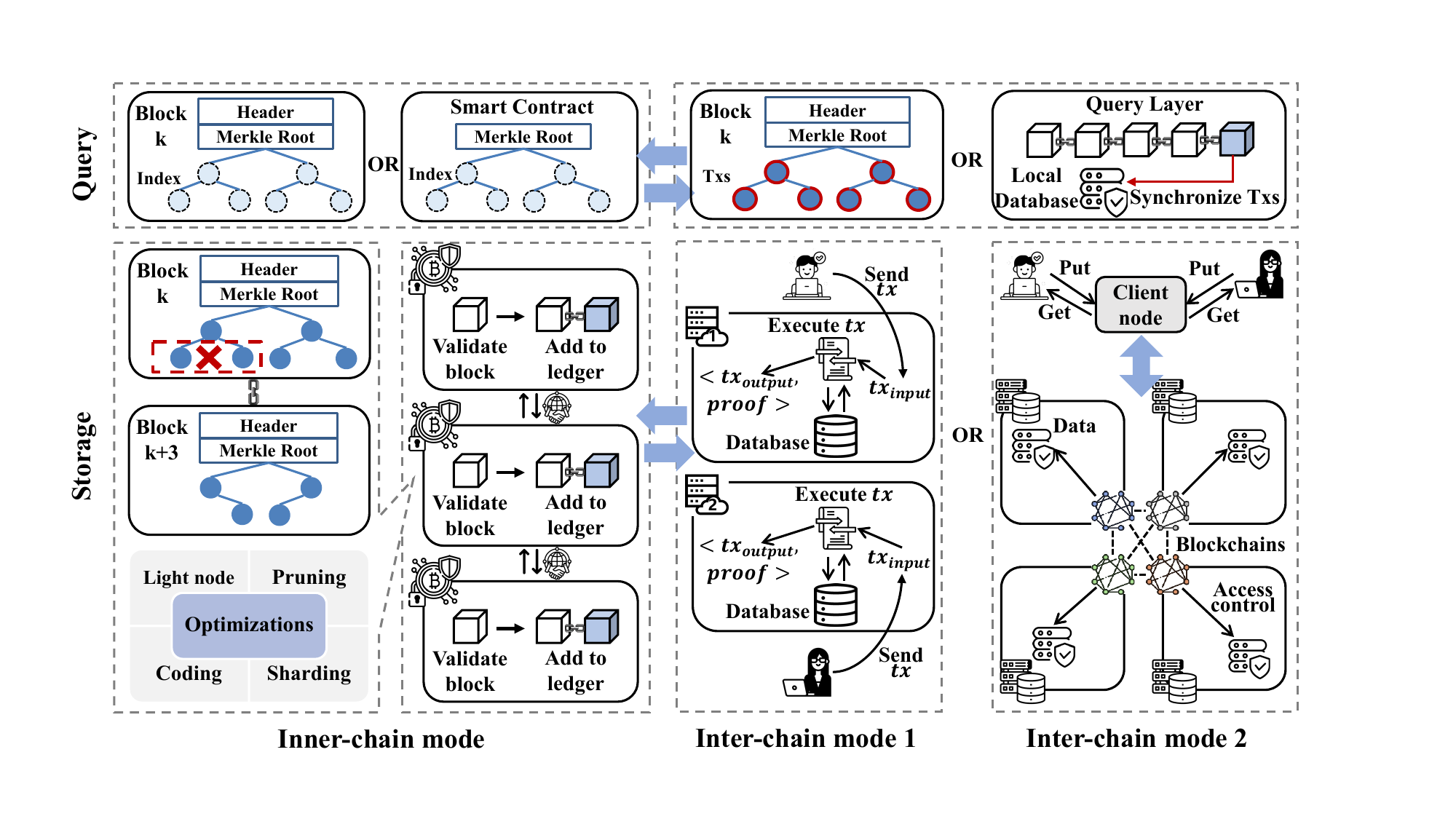}
  \caption{The overview of blockchain storage and query.}
  \label{storage_query_fig}
  \vspace{-0.2in}
\end{figure*}

\section{Data Scalability}
\label{data}
In this Section, we showcase the scalability of blockchain data from inner-chain and inter-chain perspectives. Within these two aspects, the existing work is categorized into data storage and data query. As shown in Fig. \ref{storage_query_fig}, inner-chain nodes primarily store partial data, data indexes, and hash values to reduce their resource consumption. Inter-chain nodes are responsible for storing detailed data and their data structure for transaction execution and speedy location. The following subsections will demonstrate them one by one.

\subsection{Inner-chain Solutions}

\subsubsection{Storage}
Typically, blockchain inner-chain nodes store the entire ledger data, which reduces scalability in terms of blockchain data. To this end, we survey the existing works to alleviate the amount of inner-chain node data storage. Existing solutions are based on the following aspects: use of light nodes, pruning, sharding, data encoding, and optimizing. 

Nakamoto~\cite{nakamoto2008bitcoin} divided nodes into light nodes and full nodes. The full node saves the data of the entire blockchain network, while the light node saves only the block header data of the longest PoW chain. When the light node needs to query the detailed transaction, it must request the data from the full nodes and compare it with the block header data saved by itself to verify whether the data sent by the full node is correct.

Through the pruning operation, the amount of data stored in the node can be reduced. Dai \emph{et al.}~\cite{dai2019jidar} proposed a puzzle-like data reduction method called the jidar. This method allows each node in the blockchain to store only the transactions they are interested in and the relevant Merkle branches. The complete block data can be put together like a puzzle with data fragments. Nodes can securely maintain and verify all their relevant data locally without trust assumptions. Experimental results show that jidar can reduce the storage cost of a node to about $1.03\%$ of the Bitcoin.


As a typical storage sharding solution, CUB~\cite{xu2018cub} defines the concept of the Consensus Unit (CU), which allows multiple nodes in the blockchain network to form a single unit. Nodes within the unit cooperate with each other to jointly store at least one complete blockchain ledger. To determine the optimal allocation strategy for existing blocks, this paper models the allocation problem of blocks to nodes as an NP-hard Block Allocation Optimization problem and proposes three effective heuristic algorithms to solve the static allocation problem. At the same time, the corresponding strategies are formulated to meet the needs of dynamic scenarios such as new block generation, node joining, or leaving the CU. Li \emph{et al.}~\cite{li2020multi} introduced a cluster-based multi-node collaborative storage strategy called ICIStrategy, wherein the participants in the blockchain network are divided into multiple clusters, with each cluster of nodes jointly maintaining a complete ledger. This strategy reduces the amount of data each participant needs to store, thus alleviating the storage pressure. Nodes within each cluster collaboratively store and verify blocks to reduce the storage pressure and communication overheads. Simulation results show that the ICIStrategy requires only $25\%$ of the storage space used by Rapidchain, effectively solving the storage limitation problem and improving blockchain performance. Inspired by CUB, Yin \emph{et al.}~\cite{yin2022ebsf} proposed the block storage framework, EBSF, and mathematically modeled the allocation of block data based on the characteristics of nodes (such as storage capacity, cost, and ability to respond to queries), with each block assigned to at least one node. The goal was to minimize the total cost of storing the entire blockchain ledger while meeting the query ability threshold of each block. As the optimization problem is NP-hard, the authors proposed three heuristic algorithms to solve it. Additionally, the paper extended the three methods to dynamic scenarios such as node joining and leaving, new block allocation, and old block pruning.

As an effective solution, Qi \emph{et al.}~\cite{qi2020bft} proposed a new storage engine, BFT-store, based on data encoding. It divides a block into n-2f sub-blocks and encodes them into n-coding blocks using erasure coding. Each node stores one of the coding blocks to reduce the data storage on a single node. BFT-store overturns the traditional full replication strategy. It reduces the storage consumption of each block from O(n) to O(1).  To ensure system scalability, an efficient four-phase re-encoding protocol is designed, and a multi-replica scheme is adopted to improve the read performance. BFT-store is implemented on the open-source licensed blockchain Tendermint and through experiments, its scalability, availability, and efficiency are demonstrated. Another work~\cite{qi2021byzantine} showcases BFT-Store which combines erasure coding with BFT consensus protocols and breaks the full replication strategy. The demonstration shows (i) how BFT-Store partitions data across all nodes and (ii) how BFT-Store recovers coding blocks in a Byzantine scenario.

For optimization, Ruan \emph{et al.}~\cite{ruan2019fine} proposed a traceability system called LineageChain, which can effectively capture the fine-grained sources of the blockchain. It securely stores the source and provides a simple access interface to smart contracts. In addition, LineageChain provides a new skip-list index to support efficient source query processing. The experimental results show that this traceability system has benefits for new blockchain applications, efficient queries, and smaller storage costs. CUB~\cite{xu2018cub} and EBSF~\cite{yin2022ebsf} combine sharding and optimizing technologies to distribute data to appropriate nodes for storage, thereby reducing the burden on nodes.


\begin{table*}
    \tiny
    \caption{Blockchain authenticated query approaches}
    \label{query_tab}
    \begin{tabularx}{\textwidth}{cXXXXc}
    \toprule
    Solutions & Authenticated queries & Indexes & Introduced technologies & Realization ways & Decentralization\\
    \midrule
    \cite{xu2019vchain} & Boolean range query and subscription query & Merkle balanced tree and inverted prefix tree & Accumulator, balanced tree, and skip list & Block header and body & No\\
    \cite{zhang2019gem} & Range query & GEM$^{2}$-tree and GEM$^{2*}$-tree & Merkle B-tree, and suppressed Merkle B-tree & Smart contract & No\\
    \cite{ruan2019fine} & Provenance query & Deterministic Append-Only Skip List (DASL) & Directed Acyclic Graph (DAG) and skip list & Smart contract & No\\
    \cite{xu2023stk} & kNN and range query & Merkle RK-Tree & R-Tree and bloom filter & Block header and body & No\\
    \cite{wu2021vql} & Key-value and range query & Merkle Patricia Tree & Merkle Patricia Tree & Database middleware & No\\
    \cite{xu2022qos} & Cardinality query & Merkle Cardinality Tree & bitstring and cardinality estimator & Block header and body & No\\
    \cite{peng2020falcondb} & (Range) Historical query and delta query & Database-based ADS & Traditional database and BFT consensus & Database and smart contract & Yes\\
    \cite{han2021vassago} & Cross-chain provenance query & Dependency graph & Hybrid blockchain and cross-chain TX execution & Auxiliary blockchains & Yes\\
    \cite{li2021bringing} & Key-value query &  Keyword-based indexes and witness tree & Trusted Execution Environment (TEE) & Search engine & Yes\\
    \bottomrule
    \end{tabularx}
\end{table*}

\subsubsection{Query}
To improve the functioning and efficiency of the blockchain query service, some of the initial works mainly synchronized the blockchain data and organized them in an easy-to-search form through a plug-in query layer or database, such as EtherQL~\cite{li2017etherql}, BigchainDB~\cite{mcconaghy2016bigchaindb}, FlureeDB~\cite{platz2017flureedb}, etc.
These solutions usually rely on a trusted third party for the correctness and integrity of query results.
In response to the problem that the service provider may return incorrect or incomplete query results, light node users need additional verification mechanisms to authenticate the results.
This paper presents a comprehensive analysis of the various blockchain-based verifiable query schemes and provides a comparative study, which is presented in Table \ref{query_tab}.
Moreover, the verifiable query research is mainly divided into the following categories: enrich query methods, improve query efficiency, reduce query overhead, and improve query decentralization.

For enriching query methods on the blockchain,
vChain~\cite{xu2019vchain} and Gem$^2$-Tree~\cite{zhang2019gem} have opened up the research on verifiable queries of blockchains.
vChain proposes a verifiable data structure based on accumulators, and designs two index structures from intra-block and inter-block, thereby ensuring the soundness and completeness of the boolean range query results.
%
%
This approach can ensure that light node users can still safely use blockchain data without saving the complete blockchain data.
This further reduces the requirements of the blockchain system on user resources and further increases the scalability of the blockchain system.
Gem$^2$-Tree implements a verifiable data structure based on smart contracts to provide verifiable range queries, thus avoiding modification of the underlying data structure.
Pei \emph{et al.}~\cite{pei2020efficient} proposed a verifiable semantic query solution for hybrid blockchain systems.
They proposed the Merkle Semantic Trie, which provides schemes such as keyword query, range query, fuzzy query, etc., and has good compatibility.
In order to query historical transactions, Ruan \emph{et al.} proposed LineageChain ~\cite{ruan2019fine}.
LineageChain provides a simple interface for smart contracts to query the historical data of account-based blockchain systems.
Xu \emph{et al.}~\cite{xu2023stk} proposed a new verifiable kNN and range query scheme based on a hybrid blockchain system for Spatial-Temporal-Keywords (STK) transactions.
Among them, MRK-Tree (replacing the Merkle Hash Tree of the traditional blockchain system) organizes STK transactions and supports verifiable kNN and range queries.
%
%
Zhang \emph{et al.}~\cite{zhang2021authenticated} proposed efficient new certifiable data structures, including the Suppressed Merkle inverted index and Chameleon inverted index, to realize verifiable keyword queries in hybrid blockchain systems.
SEBDB~\cite{zhu2019sebdb} designed a new type of blockchain database, which brings the advantages of traditional databases to the blockchain and improves the scalability of the blockchain.
SEBDB introduces relational semantics to blockchain transactions and supports SQL-like language for general data operations.
The multi-layer structure helps SEBDB efficiently organize blockchain data, and SEBDB supports verifiable rich queries as an extension of the blockchain's verifiable query capabilities.
LedgerDB~\cite{yang2020ledgerdb} proposed a centralized database to provide high audibility, low storage overhead, and high throughput, starting from the performance of current blockchain data audibility and real-world needs.
LedgerDB introduced the TSA two-way anchor protocol to resist the malicious behaviors of users or SPs.
Furthermore, LedgerDB provides verifiable data deletion operations to clean up obsolete or hidden data, thereby improving storage scalability.
On this basis, Yang \emph{et al.}~\cite{yang2022ubiquitous} set the three-element verification factor of what-when-who, so that the data ledger was classified in a standardized manner (Dasein-complete) for practical use in the real world.

To improve the query efficiency, LineageChain~\cite{ruan2019fine} converted the Merkle Tree into Merkle DAG and introduced the Deterministic Append-only Skip List (DASL) based on the skip table to improve query efficiency.
In order to accelerate the query efficiency, Xu \emph{et al.}~\cite{xu2023stk} designed an efficient block pruning (EBP) algorithm for multi-block querying and the Authenticated kNN/Range Query (AK/RQ) algorithm for single-block querying.
Aiming at the problem of low efficiency of large-scale blockchain data querying, Xu \emph{et al.}~\cite{xu2022qos} proposed a scheme for estimating the number of blockchain transactions.
The program could intelligently adjust the query efficiency and precision according to the user settings.
Xu \emph{et al.} proposed the MCT data structure to store bitstrings for efficient cardinality estimation.
Furthermore, they designed a DOSE algorithm to terminate the estimation protocol while guaranteeing the estimation accuracy, dynamically.
BF-DOSE further improves the efficiency of the DOSE algorithm by pruning non-target blocks.
%
%
Linoy \emph{et al.}~\cite{linoy2022authenticated} proposed a new verifiable data structure, Authenticated Multi-Version Skip List (AMVSL), to support a range query of blockchain historical data.
AMVSL supports efficient data maintenance and is identified according to the version of data.
At the same time, AMVSL can also query across multiple versions.
Based on AMVSL, Linoy \emph{et al.} implemented three types of range queries, which were used for single data version range querying, multiple data versions range querying, and multiple data versions all-key querying.

To reduce the query overhead, Gem$^2$-Tree~\cite{zhang2019gem} proposed new data structures Merkle B-Tree (MBT) and Suppressed MBT (SMBT) to reduce the consumption of queries and data updating.
On the basis of MBT and SMBT, Gem$^2$-Tree further proposed a two-layer index structure and optimized the cost of maintaining data according to the data distribution.
Although the smart contract-based approach is more flexible, the storage and computing overhead of the contract must be addressed. 
Pei \emph{et al.}~\cite{pei2020efficient} offloaded the inner-chain overhead to the inter-chain. 
The MST they proposed can be used to organize inner-chain transactions efficiently, and store information such as retrievable semantics and locations on the chain in the form of transactions.
The inter-chain part stores the original data and maintains a mapping relationship with the inner-chain data.
LVQ~\cite{dai2020lvq} proposed a verifiable query scheme based on the Bloom Filter (BF), which provides an efficient storage scheme for light nodes.
LVQ improves the Bitcoin system by adding the hash of BF to the block header, which certifies the correctness of BF.
LVQ introduces a novel data structure Bloom filter integrated Merkle Tree (BMT) for merging BF to further reduce the communication overhead of light nodes.
At the same time, LVQ's Stored Merkle Tree (SMT) data structure makes up for the false positive problem of BF and can reduce communication overhead by proving the absence of data.
VQL adds an intermediate layer (which can be provided by cloud service providers) between the blockchain system and the application layer to provide trusty querying services.
VQL also synchronizes data from underlying databases and reorganizes the data in cloud servers.
In order to ensure the verifiability of the query service, VQL uses encrypted fingerprints to verify the data in the middle layer.
The encrypted fingerprint is embedded in the blockchain system to ensure that it cannot be tampered with.
The Suppressed Merkle inverted index in~\cite{zhang2021authenticated} is beneficial for light nodes with logarithmic maintenance costs. 
The Chameleon inverted index further reduces the maintenance costs.
%

When considering the improvement of query decentralization, FalconDB~\cite{peng2020falcondb} is a blockchain-based database designed to balance the performance of shared databases, in terms of security, efficiency, and compatibility.
FalconDB tolerates $(N-1)/3$ malicious participants while ensuring that clients can check historical data and recover from malicious tampering.
FalconDB builds a verifiable data structure to support light nodes in verifying the results returned by the full nodes.
The corresponding incentive model of FalconDB promotes the positive behavior of each node in the system.
Li \emph{et al.}~\cite{li2021bringing} proposed a TEE-based decentralized search scheme, DeSearch, that simultaneously ensures query verifiability and privacy protection.
DeSearch's witness mechanism ensures the correctness and integrity of query results and reduces the computational overhead of query and proof by reusing historical queries.
DeSearch has designed a public information service that helps executors share data and agree on a data snapshot.
At the same time, DeSearch can also protect query privacy, which includes the protection of the query methods and returned data volume.

\begin{table*}
    \tiny
    \caption{Blockchain storage approaches}
    \label{storage_tab}
    \begin{tabularx}{0.85\textwidth}{ccccl}
    \toprule
    Solutions & Proposals & Mechanisms & Types & Description\\
    \midrule
    \cite{nakamoto2008bitcoin} & Bitcoin & Light nodes & Inner-chain & Keeping block header data\\
    \cite{dai2019jidar} & Jidar & Pruning & Inner-chain & Keeping interested transactions\\
    \cite{xu2018cub} &  CUB & Sharding, Optimizing & Inner-chain & Jointly storing ledger by a shard \\
    \cite{li2020multi} &  ICIStrategy & Sharding & Inner-chain & Jointly storing ledger by a shard\\
    \cite{yin2022ebsf} & EBSF & Sharding, Optimizing & Inner-chain & Jointly storing ledger by a shard\\
    \cite{qi2020bft,qi2021byzantine} & BFT-store & Encoding & Inner-chain & Encoding block into sub-blocks\\
    \cite{ruan2019fine} & LineageChain & Optimizing & Inner-chain & Allocating data to fit node\\
    \cite{xu2021slimchain} & SlimChain & Inner-chain, Inter-chain & Inter-chain & Inter-chain processing and inner-chain verification\\
    \cite{poon2016bitcoin} & Lightning Network & Inner-chain, Inter-chain & Inter-chain & Inter-chain processing and inner-chain verification\\
    \cite{kalodner2018arbitrum} & Arbitrum & Inner-chain, Inter-chain & Inter-chain & Inter-chain processing and inner-chain verification\\
    \cite{el2019blockchaindb} & BlockchainDB & Hybrid blockchain database & Inter-chain & Integrating the blockchain into the database\\
    \cite{konsta2021clouseau} & Clouseau & Hybrid blockchain database & Inter-chain & Integrating the blockchain into the database\\
    \cite{peng2020falcondb} & FalconDB & Hybrid blockchain database & Inter-chain & Integrating the blockchain into the database\\
    \bottomrule
    \end{tabularx}
\end{table*}

\subsection{Inter-chain Solutions}
\subsubsection{Storage}
Another way to improve the scalability of blockchain data is through inter-chain storage. The core idea is to store the data on third-party servers, and only store the data summary on the chain. The current research mainly falls into two categories. The first designs verification strategies based on blockchain and implements inter-chain processing and inner-chain verification of data. The second integrates the blockchain into the database, to achieve access control and data management of a hybrid distributed database. The main relevant work is presented in Table \ref{storage_tab}.

For the first type, Xu \emph{et al.}~\cite{xu2021slimchain} proposed a stateless blockchain system, SlimChain, which can scale transactions through inter-chain storage and parallel processing. The main idea is to use inter-chain storage nodes to store ledger states and simulate smart contract execution. It maintains short-term commitments to ledger states only inner-chain, and includes inter-chain smart contract execution, inner-chain transaction verification, and state commitments. SlimChain optimizes network transmission and further improves the system scalability through sharding. Zhang \emph{et al.}~\cite{zhang2021authenticated} proposed a hybrid storage model in which only a small amount of metadata is stored on the chain; the original data is outsourced to inter-chain storage service providers. A new ADS scheme for certified keyword search in hybrid storage blockchains was also proposed in the paper. This scheme maintains only a part of the ADS structure and can perform secure updating using logarithmic-sized cryptographic proofs. Experimental results show that this hybrid storage model using the new ADS scheme effectively reduces the average maintenance cost on the chain without sacrificing too much query performance. Poon \emph{et al.} proposed a decentralized system called the Lightning Network~\cite{poon2016bitcoin}. Its core idea is to send transactions, which should originally have been settled on the chain through a micro-payment channel network (payment or transaction channel) off the chain, with the value transferred outside the blockchain. The channel only needs to communicate with the Bitcoin network during "creation" and "closure", and maintains peer-to-peer communication at other times. The transaction content need not be put on the chain. When a dispute arises between the two parties in the transaction, it is arbitrated on the chain. The fairness and security of inner-chain arbitration ensure that malicious users of inter-chain transactions will not act maliciously. This way, the network can be scaled. Kalodner \emph{et al.}~\cite{kalodner2018arbitrum} designed a cryptocurrency system called Arbitrum that supports smart contracts. In Arbitrum, users can code and implement smart contracts, and run them as virtual machines (VMs). Arbitrum uses incentivization mechanisms that allow users to reach a consensus on what the VMs will do inter-chain. In other words, VMs can create and complete execution without leaking the VM execution process. Therefore, Arbitrum validators need to track only the hash value of the VM state, and not the entire state. If a party acts maliciously, the validators will identify and punish the dishonest party through a challenge-based protocol. Moving the verification of VM behavior inter-chain significantly improves the scalability and privacy.

\begin{figure*}[t]
  \centering
  \includegraphics[width=0.7\linewidth]{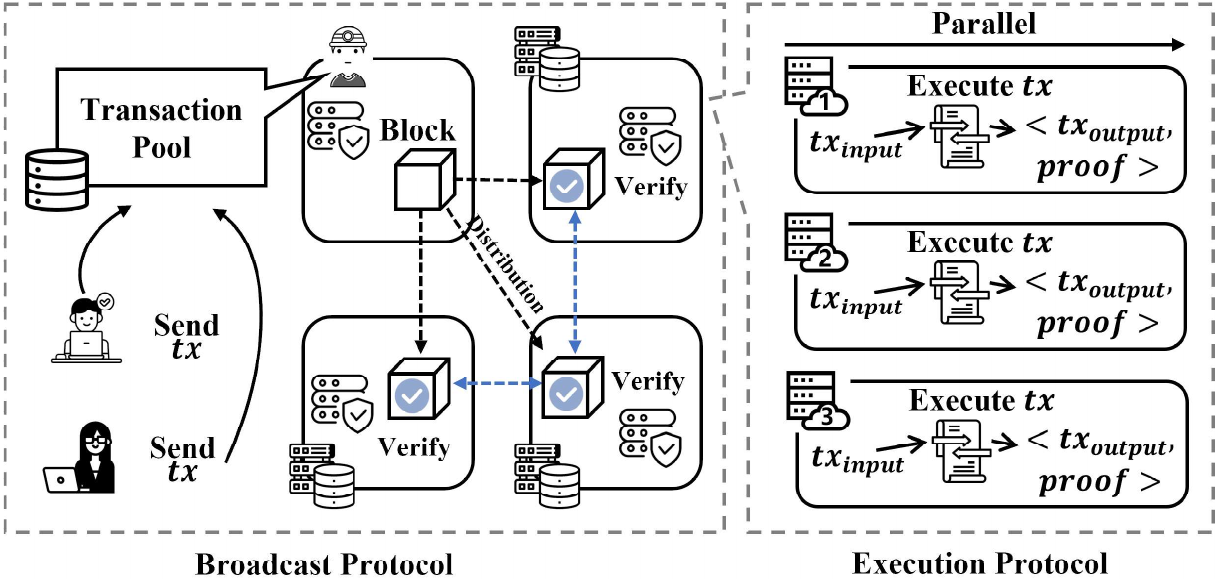}
  \caption{The process of block broadcast in blockchain network.}
  \label{broadcast_paralle}
\end{figure*}

Regarding the second type, Ge \emph{et al.}~\cite{ge2022hybrid} conducted research and qualitatively compared five existing hybrid blockchain database systems. The first system is Veritas based on Apache Kafka, which targets CFT application scenarios; the second one is Veritas based on Tendermint, which targets BFT application scenarios; the third one is BlockchainDB, which uses Ethereum as the underlying blockchain system; and the final two systems are the default version of BigchainDB which uses Tendermint and its optimized version. The default version has a blockchain pipeline function, while the optimized version adds parallel transaction verification based on the pipeline function. Experimental analyses show that Veritas (Kafka) performs better than the other systems and CFT applied to distributed databases performs better than BFT applied to blockchain-specific scenarios. El-Hindi \emph{et al.}~\cite{el2019blockchaindb} proposed a shared database on the blockchain called BlockchainDB, which uses the blockchain as the storage layer and introduces a database layer on top of it. BlockchainDB extends the blockchain through classical data management techniques and standardized query interfaces, to promote the adoption of the blockchain in data-sharing use cases. Experimental results show that BlockchainDB can provide a throughput that is two orders of magnitude higher than that of the native blockchain, improving the performance and scalability of the blockchain. HDFS is vulnerable to attacks from malicious users and participating nodes, and cannot provide a trusted lineage mechanism. As a remedy, Konsta \emph{et al.}~\cite{konsta2021clouseau} proposed Clouseau, a system that integrates HDFS with the Ethereum blockchain. The blockchain provides verifiable integrity on top of HDFS and acts as a security coordinator, supplementing the existing HDFS Namenode. In addition, to ensure performance, Clouseau maintains minimal information on the chain, so that the system will not incur significant overhead on the critical paths of read/write operations. During the system demonstration, attendees can interact with Clouseau, disrupt data, and witness how Clouseau detects malicious behavior. Grabis \emph{et al.}~\cite{grabis2020blockchain} proposed an efficient method for distributed data storage and data sharing. The main idea was to use blockchain to control access to personal data and use a knowledge base to improve retrieval efficiency. The conceptual model and data management process were elaborated, and a prototype was developed. This paper compared this prototype with inner-chain storage techniques, and experimental results showed that this approach consumed less storage space and allowed faster data retrieval. Peng \emph{et al.}~\cite{peng2020falcondb} proposed FalconDB, which enables efficient and secure collaboration in all aspects of the database in the case of limited hardware resources. FalconDB uses a database server with a validation interface accessible to the client, and stores digests on the blockchain for query/update authentication. Using blockchain as a consensus platform and distributed ledger, FalconDB can work in situations where there is mutual distrust. At the same time, FalconDB incurs minimal storage costs for each client and provides any available, real-time, and concurrent access to the database. Therefore, FalconDB ensures that individual users can participate in collaborations with high efficiency, low storage costs, and blockchain-level security guarantees.

\subsubsection{Query}
In order to further improve the scalability of the blockchain query, some works have started exploring multi-chain queries.
Han \emph{et al.}~\cite{han2021vassago} proposed the Vassago system to achieve fast traceability of cross-chain transactions.
The basic idea of Vassago is to save the dependencies of cross-chain transactions and then perform further queries based on the dependencies.
Vassago consists of a two-tier architecture, including the dependency blockchain and transaction blockchain, which store dependency and transaction information, respectively. 
The transaction dependencies ensure the verifiability of the query results and provide the possibility of executing query tasks in parallel.
Qanaat~\cite{amiri2022qanaat} is a multi-enterprise-oriented blockchain system that ensures the privacy and security sharing of multi-enterprise business data.
Qanaat designs layered data models and stores data collections separately.
Data collections are shared only when necessary (enterprise collaboration), which is also the smallest subset of data.
To ensure data consistency, including local and global data consistency, Qanaat builds a DAG-based data structure for each enterprise.
Such a scheme efficiently organizes internal and cross-enterprise data and improves the scalability of the blockchain on the basis of protecting data confidentiality.

\begin{table*}
    \caption{Blocckhain Propagation}
    \tiny
    \label{broadcast}
    \begin{tabularx}{\textwidth}{cccX}
    \toprule
    Solutions & Optimization Method & Protocol & Performance\\
    \midrule
    \cite{zhang2021accelerating}& Reputation & Block Propagation Protocol & {Accelerate the data propagation and optimize the usage of nodes' bandwidths} \\
    \cite{zhao2022bodyless} & Pre-verification & Block Propagation Protocol & {Reduce propagation time and TPS is limited by the node hardware performance} \\ 
    \cite{ayinala2020pichu} & Pipeline & Block Propagation Protocol & {Reduces the mining interval and increases throughput} \\
    \cite{wang2022data} & Optimization Algorithm & Block Propagation Protocol, Network Topology & {Reduce propagation latency and generate the throughput} \\
    \cite{bagaria2019prism} & Decouple & Consensus Protocol & {Confirmation latency for honest transactions proportional to the propagation delay, with confirmation error probability exponentially small in the bandwidth-delay product} \\
    \cite{chen2021reducing} & Trust Roles & Block Verification Protocol & {Reduce the propagation delay and reduce the blockchain forks} \\
    \cite{zhang2021speeding} & Encoding & Block Propagation Protocol & {Increase TPS while maintaining the same propagation delay as the conventional block propagation protocols} \\
    \cite{hu2022dino} & Information Compression & Block Propagation Protocol & {Scale well with high transaction generation rates and reduce block propagation latency} \\
    \cite{bi2018accelerated} & Node Features & Block Propagation Protocol & {Decrease the average propagation latency and maximum propagation latency}  \\
    \bottomrule
    \end{tabularx}
\end{table*}

\section{Protocol Scalability}
\label{protocol}

Blockchains involve multiple protocols during operation. 
In a blockchain system, as shown in Fig. \ref{broadcast_paralle}, the broadcasting protocol is responsible for the transmission of information between nodes. 
The scalability of the broadcasting protocol has a significant effect on the reliability and performance of the blockchain system. 
If the broadcasting protocol fails to meet the demands of high concurrency and large-scale transactions, it can affect the efficiency and reliability of transactions. 
Transaction execution refers to the process where each node updates new transaction data in its local ledger. 
The speed and efficiency of transaction execution are important factors in the scalability of the blockchain system, especially under high load and frequency.

The scalability of inter-chain protocols actually refers to the interoperability of cross-chain protocols.
The scalability of the protocol, built upon data scalability, further provides performance, liquidity, and security support for architecture scaling.
In the end, they collectively ensure the scalability of the entire blockchain system ecosystem.
In the process of interoperation between blockchains, the notary mechanism ensures the reliability and tamper-proofing of cross-chain transactions by introducing multiple notaries. 
The notary mechanism can enhance the stability of cross-chain interaction and provide reliable support for cross-chain systems. 
The payment channel technology facilitates high-frequency and small-value transactions by building point-to-point channels between blockchains. 
The payment channels can reduce the confirmation time and transaction cost of cross-chain transactions and significantly improve the transaction speed and scalability. 
Atomic swaps can ensure the safety of exchanges between two chains and effectively reduce the transaction cost in the intermediate process. 
Similar to payment channel technology, atomic swaps can reduce the confirmation time and transaction cost of cross-chain transactions while enhancing the scalability of cross-chain interaction.

\subsection{Inner-chain Solutions}
\subsubsection{Propagation Protocols}

Using more efficient broadcasting protocols can prevent malicious attacks, and improve the system's trustworthiness and scalability~\cite{heilman2015eclipse}. Therefore, optimizing broadcasting protocols is an important direction to enhance the scalability of blockchain systems. As a typical distributed network, blockchain architecture creates a significant amount of communication between its nodes. Communication mainly occurs through two methods: network messaging, which enables the nodes in the network to achieve consensus, and block delivery, which is critical for competitive blockchain platforms like Bitcoin that rely on faster block delivery to gain a competitive advantage. In blockchain systems, the efficiency of network message transmission is a key determinant of performance. Researchers have proposed many methods to speed up the spread of blockchain networks, as shown in Table \ref{broadcast}. These methods, including block compression, optimizing broadcasting protocols, and using intermediaries and reputation values, aim to reduce the time taken for messages to be confirmed and the network bandwidth they use. This can make the blockchain network more effective and efficient.

Hu \emph{et al.}~\cite{hu2022dino} introduced a novel peer-to-peer block transmission mechanism for inter-node communication in the blockchain network. This approach leveraged Dino blocks that, when received by a network node, allow the node to recover the original block by utilizing transactions from the node's local transaction pool, thereby reducing the block's network transmission capacity requirements and propagation time. Additionally, Dino communicates block-building rules, instead of compressed block data, thus offering greater scalability to handle blocks containing a large number of transactions. 
In competitive blockchain systems, such as Bitcoin, efficient block propagation is critical to performance. The ability to propagate blocks quickly determines the effectiveness of block out in such systems. Bi \emph{et al.}~\cite{bi2018accelerated} suggested a methodology for selecting the nearest neighbor nodes in a blockchain network by estimating the transmission delay between nodes. By selecting the broadcasting node based on the message transmission distance, the messages could be efficiently transmitted throughout the network in a timely manner. 
Zhang \emph{et al.}~\cite{zhang2021speeding} proposed a new approach to boost the throughput of the Bitcoin system without altering the core consensus protocol components. The proposed technique involves replacing the store-and-forward relaying system with a more efficient cut-through strategy and improving the block propagation efficiency during transmission by utilizing erasure code techniques. The most significant advantage of this protocol is its ability to enhance the performance of Bitcoin without making any alterations to the data structure or cryptographic functional components of the system. Consequently, the proposed protocol can be easily integrated into the existing Bitcoin blockchain. 
In an attempt to enhance the scalability of blockchain networks, Chen \emph{et al.}~\cite{chen2021reducing} proposed the GVScheme, which introduces a guarantor function to ensure that blocks are propagated in a reliable manner. When a node receives a block from a guarantor, the order of block confirmation and propagation is determined by the trust value of the guarantor. Thus, this technique minimizes both block validation and propagation delays in the network. As noted in~\cite{zhang2021speeding}, the proposed strategy requires minimal modifications to the existing protocol and can be seamlessly integrated into the current blockchain networks. 
In distributed systems, communication bandwidth and propagation latency are usually the two main physical network properties that make it difficult to use blockchain protocols. Prism, a new blockchain technology, that is dependent on PoW, was introduced by Bagaria \emph{et al.}~\cite{bagaria2019prism}. It maximizes the physical bandwidth consumption and optimizes system performance by utilizing a structured DAG model. The DAG model separates the consensus process into different blocks and arranges them based on their roles. Prism optimizes the system performance and maximizes the physical bandwidth consumption.

According to Wang \emph{et al.} \cite{wang2022data}, enhancing broadcasting performance can significantly improve the performance of blockchain systems from the perspective of blockchain broadcasting protocols. Unfortunately, the current broadcast protocols in blockchain, such as Gossip and distributed hash table, fail to meet the requirements of low redundancy and low propagation delay. As a result, they proposed a new broadcasting mechanism, named Swift, which optimizes the P2P topology building and broadcasting algorithm in structured networks through unsupervised learning and greedy algorithms. Swift efficiently minimizes the propagation latency of blockchain P2P networks while reducing the waste of redundant bandwidth. 
Ayinala \emph{et al.} \cite{ayinala2020pichu} proposed the PiChu method for improving the scalability of blockchain networks by accelerating block propagation through pipeline technique and the design of verification blocks. The acceleration of block propagation decreases the probability of mining intervals and forks, leading to an increase in throughput. The PiChu architecture can be applied immediately to the existing blockchain networks.

Zhao \emph{et al.} \cite{zhao2022bodyless} introduced a transaction selection, sorting, and synchronization algorithm that accelerates consensus among nodes. However, transactions that rely on the coinbase address, cannot be pre-executed or pre-verified because the coinbase address of the next block miner is unpredictable. The authors proposed an algorithm to handle unresolvable transactions to attain a consistent and high TPS scheme. This scheme adopted a transmission process similar to that of PiChu, wherein most transactions are not required to be verified and transmitted during block propagation, removing the dependence of propagation time on the number of transactions in a block, and fully enabling the system to be TPS scalable.
Zhang \emph{et al.} \cite{zhang2021accelerating} introduced the concept of reputation and proposed a unique relaying protocol called RepuLay to accelerate the transmission of network transactions. The reputation system was intended to aid nodes in identifying unreliable and inactive neighbors. Each node maintains a local list of all its neighbors' reputations and processes transactions using a probabilistic technique that relies on a reputation mechanism. More precisely, a relay node examines a transaction with a defined probability, upon receiving it. Subsequently, the relay node transmits both legal and unvalidated transactions to multiple neighbors, with each neighbor having a chance of being selected as a recipient.

\subsubsection{Transaction Parallelism}
\begin{table*}
    \caption{Blockchain Transaction Parallelism}
    \tiny
    \label{parallel}
    \begin{tabularx}{\textwidth}{cXXXXX}
    \toprule
    Solutions & Applicable scenarios & Optimization Method & Concurrency control & Goal \\
    \midrule
    \cite{xiao2022nezha} & DAG-based blockchain & Graph analysis & - & {Increase the effective throughput and decrease the transaction processing latency}\\
    \cite{jin2021high} & Smart contracts in permissioned blockchain & Graph-segmentation & OCC & {Improves the execution efficiency of primary and validators, reduces communication overload} \\ 
    \cite{ponnapalli2021rainblock} & Permissionless blockchain & Removing the $I/O$ bottleneck & - & {Speedup the transaction execution} \\
    \cite{xu2021slimchain} & Permissionless blockchain & Inter-chain and sharding & OCC or SSI & {Reduce the inner-chain storage requirements and improve the throughput} \\
    \cite{amiri2019parblockchain} & permissioned blockchain & Graph analysis & Dependency graphs & {Improve the performance on both order-execute and execute-order blockchain} \\
    \cite{garamvolgyi2022utilizing} & Smart contract on permissionless blockchain & Graph analysis & OCC & {Achieve the high speed of execution}  \\
    \cite{dickerson2017adding} & Smart contract on permissionless blockchain & Fork-join & Reordering & {Increase the throughput and provide better control of concurrency} \\
    \cite{wust2020ace} & Smart contract on permissionless blockchain & Inter-chain & - & {Enables several orders of magnitude more complex smart contracts than standard Ethereum} \\
    \cite{reijsbergen2020exploiting} & Permissionless blockchain & Graph analysis & Dependency Graph & {Improve transaction execution efficiency in Ethereum} \\
    \bottomrule
    \end{tabularx}
\end{table*}


Transaction execution is of vital importance in the operation of blockchain systems. Slow transaction speeds may severely limit the system's scalability. To address this issue and improve transaction execution speed and system scalability, researchers have proposed a combination of parallel and concurrent techniques. Furthermore, to mitigate the issue of high concurrent transaction conflict rates in smart contract scenarios, relevant research has incorporated concurrency control and graph analyses to avoid transaction conflicts while enhancing parallel execution efficiency.
There are mainly two cases of transaction execution model, order-execute model, and execute-order model. Various solutions have been proposed in related studies to address the problem of transaction parallelism, as shown in Table \ref{parallel}. 

Daniël \emph{et al.} \cite{reijsbergen2020exploiting} contended that apart from consensus, transaction execution is the second-most critical module that influences blockchain performance and security. The authors collected historical data from seven blockchain systems, including Ethereum, Bitcoin, and Zilliqa, and analyzed them using two metrics (\emph{single-transaction conflict rate per block} and \emph{group conflict rate per block}). They found that UTXO-based blockchains have more concurrency than account-based ones. Analytical models were proposed for single transaction concurrent execution and group concurrent execution to estimate the transaction execution speed for a given level of concurrency. The models were validated on the seven blockchain systems.

Asynchronous and Concurrent Execution of Complex Smart Contracts (ACE) was developed by Wüst \emph{et al.} \cite{wust2020ace} with the goal of enabling complex smart contract execution on permissionless blockchains through an improved concurrency control mechanism and flexible trust model. ACE employs an inter-chain execution method whereby the contract creator specifies a group of service providers to independently execute the contract code, separate from the consensus layer. ACE distinguishes itself from prior solutions as it enables secure smart contract initiation of contract execution across different service providers whilst allowing secure concurrency control. ACE is the first of its kind to exhibit the capability of supporting the inter-chain execution of interactive smart contracts with flexible trust assumptions.

Dickerson \emph{et al.} \cite{dickerson2017adding} introduced a new technique by which miners and verifiers could run smart contracts that did not conflict, in parallel. This method used a deterministic \emph{fork-join} \cite{blumofe1995cilk} program that stores a serializable concurrent scheduling sequence. The verifiers use the scheduling sequence to execute and verify the contracts.
Garamvölgyi \emph{et al.} \cite{garamvolgyi2022utilizing} conducted a thorough analysis of the historical transaction execution in Ethereum and found that smart contracts often face obstacles in achieving concurrent execution. To overcome these obstacles, they proposed a conflict resolution technique that involves using partition counters and swappable instructions. This approach can enhance the execution speed of contract transactions. Furthermore, they introduce a new scheduling scheme, OCC-DA, which is an optimistic concurrency control scheduler with deterministic aborts, designed to enable the use of OCC scheduling in permissionless blockchains. 
Amiri \emph{et al.} \cite{amiri2019parblockchain} contended that most existing blockchains are inadequate in addressing the potential issues of distributed system applications and have serious architectural limitations. To address these concerns, they proposed the OXII paradigm, which is an approach that allows concurrency control of smart contract transactions by constructing a dependency graph to identify transaction conflicts and determine the order of execution. This strategy allows permissioned blockchains to support concurrent transaction execution. They also proposed the ParBlockchain prototype under the OXII paradigm, which was experimentally verified to be suitable for smart contract transaction scenarios with varying levels of competition.

SlimChain \cite{xu2021slimchain} utilizes the concept of inter-chain parallel execution and inner-chain state confirmation. This approach moves transactions inter-chain to be executed in parallel, while also ensuring secure inter-chain execution through the use of TEE. To address the challenge of arbitrary commit orders, SlimChain applies OCC along with Serializable Snapshot Isolation (SSI), employing the heuristic approach presented in \cite{cahill2009serializable} to achieve efficient concurrency control.
RainBlock \cite{ponnapalli2021rainblock} improves the performance of public blockchains without changing the original PoW consensus logic by eliminating the Input/Output (I/O) bottleneck in transaction processing, which enables miners to process more transactions simultaneously. The main contributions of RainBlock are twofold: 1) the proposal of the RainBlock architecture to eliminate I/O from the critical path of transaction processing, and 2) the design of a distributed multi-version DSM-Tree-based data structure that efficiently stores the system state.

Chen \emph{et al.} \cite{chen2021peep} identified two challenges associated with transaction parallelism in the existing blockchain systems: 1) differences in the order of concurrent execution across different nodes, and 2) the inability of the state tree to support efficient concurrent updates. To address these challenges, they propose the Parallel Execution Engine Protocol (PEEP), which utilizes a deterministic concurrency mechanism for parallel execution with a predefined serial execution order for fetches. PEEP provides parallel update operations for the state tree and can guarantee compatibility with various Merkle tree-based state trees. 
Jin \emph{et al.} \cite{jin2021high} introduced a novel two-stage concurrency control protocol that optimized the two-stage-style concurrent execution process of smart contracts. To execute transactions, the system generated a transaction-dependent graph with high parallelism for the verifier and designed a graph partitioning algorithm to split the graph into several subgraphs. This maintains parallelism and substantially reduces communication costs. Additionally, a deterministic replay protocol was proposed to facilitate faster concurrent scheduling. Integration of the proposed two-stage protocol with the Practical Byzantine Fault Tolerance (PBFT) was suggested to further improve optimization.

Xiao \emph{et al.}~\cite{xiao2022nezha} aimed to enhance the system throughput and decrease processing latency by investigating address dependencies among transactions. To achieve this goal, they proposed an efficient DAG-based blockchain concurrency control scheme, named NEZHA. NEZHA intelligently builds an address-based conflict graph (ACG), using address dependencies as edges, to capture all conflicting transactions. The authors apply a hierarchical ranking algorithm to generate a total order among transactions by ranking the transactions on each address based on the ACG and derived ranking hierarchy.

\subsection{Inter-chain Solutions}
\subsubsection{Notary}

As an \textit{inter-chain} scaling protocol, the notary mechanism can increase the cross-chain contact stability and offer trustworthy support for the cross-chain system. 
The notary mechanism is classified into two approaches: centralized notary and decentralized notary. 
Designed by Ripple~\cite{hope2016interledger}, the Interledger Protocol enables two distinct ledger systems to seamlessly exchange currencies with one another via an intermediary known as a "connector," which, in practice, operates as a centralized notary. 
This protocol eliminates the need for trust between the parties involved in the transaction, and importantly, ensures that the connector neither loses nor misappropriates funds.
As for centralized notary, PalletOne~\cite{palletone} supports multiple chains in smart contracts, through jury consensus and adapters to operate on different blockchains, eliminating the need for parallel chains. 
Users can utilize PalletOne passes as transaction fees to incentivize the jury in driving the PalletOne technology. 
UniswapV3 \cite{adams2021uniswap} is an unregulated, automated market-making protocol built on the Ethereum blockchain.
It overcomes the naturally low metallic profitability of constant-function market makers, improves the accuracy and convenience of price forecasting, and provides a more flexible fee structure.

On the contrary, Corda \cite{hearn2016corda} designed a highly available notary cluster that could include multiple worker nodes distributed to multiple data centers, with a database cluster on the back end to hold transaction information. 
This notary cluster as a whole provides services to the public. 
As a decentralized notary, it is composed of multiple working nodes that form a distributed architecture, which elects master nodes to provide services and ensure data consistency through a distributed consensus mechanism. 
Tokrex~\cite{mayer2017tokrex} brings a completely decentralized approach to the interoperability of blockchain systems.
It is a meta-system that enables the exchange of assets between different blockchains (cross-chain) as well as within a blockchain (intra-chain) in a real-time setting. 
0x~\cite{warren20170x} designed a protocol that facilitated low-friction peer-to-peer exchange of ERC20 tokens on Ethereum.
It was intended to serve as an open standard and common building block, driving interoperability among decentralized applications that incorporate exchange functionality. 
An intermediary role in the 0x protocol, called a relayer, helps broadcast orders and can choose to charge a fee per facilitated transaction. 

\subsubsection{Payment Channel}

Payment channel technology, also known as micropayment channels, reduces transaction fees and improves system throughput by establishing \textit{inter-chain} peer-to-peer channels to aggregate small transactions with high frequency. 
The concept of payment channels was introduced by LN~\cite{poon2016bitcoin} as a decentralized system where transactions are sent through a network of micropayment channels. 

Raiden Network~\cite{hees2016raiden} is an implementation of payment channel technology specifically designed for Ethereum.  
The Raiden Network preserves the security mechanism that the blockchain system has through peer-to-peer payments and margin deposits in the Ethernet network. 
Raiden nodes interact with Ether nodes to facilitate transfers and communicate with other Raiden nodes, as well as with the Ethereum blockchain for managing margin deposits. 

From the perspective of performance and resource overhead, Guo \emph{et al.}~\cite{guo2019measurement} conducted a comprehensive evaluation of newly proposed protocols aimed at enhancing the performance of Lightning Networks (LNs) based on data collected over a 15-month period. 
The study focused on analyzing the success rate and dispersion of payment routing. 
The analysis encompassed the success rate of payment routing, and the level of decentralization, and provided a comprehensive understanding of network mechanisms.
To mitigate the increased consumption of inner-chain resources caused by the gradual exhaustion of a substantial portion of payment channels within payment channel networks, Xu \emph{et al.}~\cite{xu2021privacy} introduced OPRE, an optimal inter-chain recovery protocol for payment channels. 
The designed protocol includes privacy-preserving features to address user privacy concerns, ensuring that the user's balance information remains undisclosed. 
The protocol achieved optimal restoration of payment channels while ensuring robust privacy guarantees utilizing cryptography.
Seo \emph{et al.}~\cite{seo2022enhancing} proposed a two-layer structured aggregated payment request scheme to extend the bandwidth in response to the limited scalability provided by the LN, i.e., constrained by channel mobility and payment request bandwidth (currently 483 in each direction).
Wu \emph{et al.}~\cite{wu2020local,9763398} introduced the notion of supernodes and the supernodes-based pooling to enhance the scalability of micropayments within a large Lightning Network (LN).
The supernodes, along with a subset of their neighboring non-super nodes, pool together to facilitate network partitioning within the LN.
To enhance the scalability of micropayments, the set of involved nodes is reduced, with only supernodes taking part in the search for and payment to other supernodes.

From a security perspective, Malavolta \emph{et al.}~\cite{malavolta2018anonymous} introduced a new attack on the existing payment channel network called a wormhole attack.
They also proposed a new encryption structure called the anonymous multi-hop lock (AMHL). 
It started from the security analysis of the existing payment channel network and reported a new attack applicable to all major payment channel networks, which allows attackers to steal fees from honest middlemen along the way. 
Additionally, the Lightning Network (LN) developers have implemented the authors' ECDSA-based AMHL in their payment channel network, thereby exemplifying the practicality, security, and privacy of this method in contemporary cryptocurrencies. 
Furthermore, the team conducted a performance evaluation using a commercial machine, wherein experiments demonstrated the strong practicability of AMHL, with all operations completed in under 100 ms and introducing a communication overhead of less than 500 bytes.
In a separate study, Kappos \emph{et al.}~\cite{kappos2021empirical} provided a thorough analysis of the privacy of the LN and analyzed several attacks that exposed privacy, including the number of tokens owned by nodes and the recipients and payers in the state channel.
Biryukov \emph{et al.}~\cite{biryukov2022analysis} developed a precise probing model that accounts for parallel channels, enabling comprehensive balance information extraction in multi-channel hops. 
The model also quantifies the information gained by attackers and proposes an optimized algorithm for selecting probe amounts in multi-channel hops.
This paper showcases the efficiency of their approach using real-world data obtained from their own LN simulator focused on probing.

\subsubsection{Atomic Swap}

The atomic swap protocol was originally conceived to facilitate asset exchanges between distinct blockchain networks. 
Its significance has grown substantially in the realm of protocol design for \textit{inter-chain} scaling, owing to its inherent attributes.
Indeed, cross-chain atomic swaps are the result of a fusion of cryptographic technology, smart contracts, and specialized role design. 
These swaps encompass several pivotal elements, encompassing incentive mechanisms, security considerations, and formalization aspects.
Herlihy \emph{et al.}~\cite{herlihy2018atomic} pioneered the introduction of the atomic cross-chain asset swap protocol.
It constructs an interactive directed graph with designated leading nodes and employs hash time lock contracts.
A hash lock can only be unlocked if not timed out, with the provided secret "s," and sequential signatures from all nodes along the path to the leader. 
Asset transfer is deemed complete when all hash locks unlock; 
otherwise, assets are returned. However, this protocol assumes fixed, known elements, such as the exchange graph, leader node, and hash lock details, limiting flexibility.
Subsequently, the atomic swap protocol underwent continuous optimization in the context of serving \textit{inter-chain} scalability.

From a design and performance optimization perspective,
Zamyatin \emph{et al.}~\cite{zamyatin2019xclaim} conceived a trustless and efficient cross-chain atomic transaction framework, XCLAIM, along with its formal definition. 
This framework aims to tackle existing issues associated with slow, inefficient, and costly cross-chain atomic transactions. 
It facilitates cost-effective token exchange between Bitcoin and Ether, leveraging self-designed cryptocurrency-backed assets.
Furthermore, it offers flexibility for migration to other established systems and their cross-chain applications.
Zakhary \emph{et al.}~\cite{zakhary2020atomiccommitment} introduced AC3WN, the first decentralized all-or-nothing atomic cross-chain commitment protocol. 
This protocol achieves atomicity and commitment in AC2T by employing cryptographic commitment schemes based on hash locks for smart contract exchange and refund. 
It ensures that all smart contracts either execute entirely or result in full refunds. 
Notably, this is accomplished through the utilization of a decentralized witness network to coordinate AC2T, thus addressing the vulnerabilities of centralized solutions.
Thyagarajan \emph{et al.}~\cite{thyagarajan2022universal} devised a universal cross-chain atomic exchange protocol that enables the secure exchange of tokens between any target and source chains by relying only on transaction signature verification without resorting to any scripting language.
The protocol also supports secure exchange of tokens between multiple parties.
Tao \emph{et al.}~\cite{tao2023atomicity} presented a new mechanism called Unity, which ensures the atomicity and confidentiality of cross-chain transactions in the event of read or write failures by utilizing permission-controlled blockchains.
Specifically, when data is not the latest version, the 4PC protocol is employed to guarantee the confirmation or abort of cross-chain transactions. 
When data is the latest version, enforcement of transactions is achieved using SSC-based smart contracts.
Glabbeek \emph{et al.}~\cite{vanglabbeek2023crosschain} introduced a cross-chain payment protocol with guaranteed success by employing the formal specification of Asynchronous Timed Automata Networks (ANTA). 
This approach is highly motivated as ensuring the success of payments is crucial for the reliability of cross-chain transactions.
Xue \emph{et al.}~\cite{xue2023Fault} combined two alternative protocols aimed at creating more expressive and fault-tolerant cross-chain exchanges. 
These protocols enable participants to propose multiple swaps simultaneously and complete a portion of them based on their individual requirements. 
Participants express their needs using predicates, with each predicate capturing acceptable payout conditions for each participant. 
The authors constructed redundant payment paths in a multi-path routing scheme, allowing for tolerance of deviations and failures among participants while ensuring the reliability and security of transactions.
Imoto \emph{et al.}~\cite{imoto2019atomic} designed and implemented a new cross-chain atomic transaction protocol with the help of signature information and improved the space complexity and the local time complexity.
Lys \emph{et al.}~\cite{lys2021r} proposed a new protocol, R-SWAP, formalized for relays and adapters. The correctness of R-SWAP was demonstrated, and its performance in terms of cost and latency was analytically evaluated. The atomic exchange between Ethereum and Bitcoin and between Ethereum and Tendermint was implemented.

From the standpoint of protocol and security analysis,
Herlihy \emph{et al.}~\cite{herlihy2022adversarial} introduced the concept and practical implementation of cross-chain transactions as a solution to managing assets in complex distributed computing environments, particularly in adversarial business scenarios.
Additionally, the paper proposes a proof mechanism utilizing BFT consensus and Proof of Work consensus. 
However, it's worth noting that the paper does not provide explicit experimental results or metrics but rather offers an overview of the methodology and principles.
Pillai \emph{et al.}~\cite{pillai2021burntoclaim} presented the Burn-to-Claim protocol, which leverages a three-phase Proof-of-Burn protocol for asset transfer and interoperability.
It achieves asset transfer by generating transfer proofs on the source network and verifying proofs on the target network. 
It has been empirically demonstrated that this method incurs lower computational costs and is integrated into the core blockchain protocol.
Xu \emph{et al.}~\cite{xu2021game} proposed a game-theoretic model to study the strategic behavior of agents implementing cross-chain atomic trading based on HTLCs by representing the success rate of the transaction as a function of variables such as exchange rates, token prices, and their volatility. 
It is found that collateralized deposits and agents dynamically adjusting exchange rates can improve transaction success rates.
Manevich \emph{et al.}~\cite{manevich2022cross} introduced the "MPC in the head" technique into the cross-chain atomic exchange protocol, implementing a new cross-chain atomic exchange protocol that operated without the concept of global time and could be terminated by both parties at any time, further testing the practical performance of this zero-knowledge proof protocol.
A game-theoretic analysis MAD-HTLC is used by Tsabary \emph{et al.}~\cite{tsabary2021mad} to demonstrate the security and to analyze its overhead by instantiating it on running blockchains of Bitcoin and Ether.
Furthermore, the study explores the potential for miners to serve as the primary enforcers by modifying the standard Bitcoin client.
Li \emph{et al.}~\cite{li2022zerocross} designed ZeroCross, a privacy-preserving cross-chain solution based on sidechains, designed to address issues such as the need for multiple payers to make simultaneous payments or fixed transaction amounts. 
Leveraging sidechain mechanisms and state-of-the-art zero-knowledge proof protocols, this paper ensures the correctness of exchanges and protects transaction privacy. 
It also designs key exchange and verification mechanisms to achieve fairness and confidentiality.
\section{Discussion}
\label{discussion}
\subsection{Architecture Scalability}
\paragraph{Inner-chain.}
Expanding the architecture of blockchain is a crucial research direction
 in the blockchain domain. 
On one hand, sharding involves cross-shard interaction and communication mechanisms, where researchers can explore efficient and secure ways between different shards. This will encompass research areas such as executing smart contracts across shards and transferring data across shards to ensure the overall consistency and integrity of the blockchain. On the other hand, considering the integration of sharding technology and DAG structures, leveraging the advantages of both to construct a more efficient and secure blockchain architecture. This integration may bring about entirely new possibilities for future blockchain systems. 

\paragraph{Inter-chain.}
As one of the most prominent blockchain technologies currently, BoBs achieve the scaling of blockchain architecture through \textit{inter-chain} interoperability. 
Firstly, the scalability of BoBs is severely constrained by security considerations. 
For instance, while the IBC of Cosmos is designed to be more flexible than the XCMP of Polkadot, with each zone capable of independent validation, it also entails higher security risks than Polkadot.
Secondly, there is a notable absence of systematic research regarding the number of honest nodes and validators in BoBs. 
This significantly hinders BoBs' scaling into the field of \textit{inter-chain} information transmission.
Furthermore, differences in smart contract languages and execution environments in \textit{inter-chain} scenarios make it challenging to achieve smart contract state migration. 
This lack of generality affects the applicability of BoBs.

\subsection{Data Scalability}
\paragraph{Inner-chain.}
The capacity of blockchain network nodes is limited. Methods such as block pruning and cooperative storage can indeed alleviate the storage burden on nodes, but this can lead to an increase in query cost. Therefore, in future research on inner-chain data, a trade-off between storage cost and query cost needs to be considered. By designing more efficient storage strategies and query structures, the goal is to achieve storage effectiveness within the tolerance of query cost. Additionally, considering that the participants in a blockchain network are people, there are bound to be social attributes and relationships among users. Transferring knowledge from the field of social networks to the blockchain may provide clever solutions to some challenging problems.
\paragraph{Inter-chain.}
Accessing data inter-chain can indeed greatly reduce the storage cost of blockchain nodes, but it increases the risk of data security. For inner-chain nodes, it is impossible to ensure the security and integrity of the data they have not stored, especially when the data is stored inter-chain. Therefore, in the research on inter-chain data, it is necessary to pay more attention to the security and integrity of the data. This requires the introduction of cryptographic knowledge to guarantee that the data stored inter-chain cannot be tampered with. When nodes retrieve data, proof of data integrity should be provided to ensure that the retrieved content is complete. Additionally, the introduction of privacy computing can ensure the privacy and security of users, making the data storage and retrieval process of blockchain more secure.

\subsection{Protocol Scalability}
\paragraph{Inner-chain.}
Enhancing the performance and scalability of blockchain systems relies significantly on the optimized parallel execution of transactions and propagation protocols.
Future research could focus on the following areas.
In terms of parallel execution, increasing the degree of parallelism in parallel execution adds complexity to maintaining database consistency, making the resolution of read/write conflicts, and ensuring consistency for concurrent transactions of utmost importance.
Additionally, parallel execution of transactions can introduce uncertainty and challenges in ensuring the consistency and correctness of transaction results.
Thus, it is imperative to research effective methods for accurately and sequentially processing concurrently executed transactions.
For propagation protocols, the current solution still faces challenges such as excessive bandwidth usage and high latency.
Future research efforts should focus on optimizing the broadcasting protocol in terms of transmission methods, content, and other aspects.
These optimizations can help reduce the network transmission burden and enhance the speed and reliability of transaction broadcasting.

\paragraph{Inter-chain.}
Although atomic swap protocols have emerged as a pivotal technology for expanding \textit{inter-chain} functionality due to their trustlessness and practicality, they continue to confront a series of challenges.
From the perspective of capital flow, extant atomic swap protocols exhibit inherent vulnerabilities stemming from the design of escrow contracts. 
This vulnerability leads to a pronounced risk of fund immobilization and transaction unfairness. 
A promising research avenue involves the exploration of non-interactive cryptographic techniques to ensure the high liquidity of capital and low latency in transactions.
From the standpoint of malicious behavior tolerance, the current research has less engagement. 
Achieving an elevated transaction success rate while preserving absolute atomicity represents a vital topic.
As for scalability, the current atomic swap protocols predominantly center on pairwise exchanges between two parties, significantly limiting the inherent scalability of these protocols. 
Devising equitable and efficient multi-party atomic swap protocols becomes an intriguing challenge. 
The combined potential of distributed signatures and multi-party secret sharing offers a compelling avenue for resolution.

\section{Conclusion}
\label{conclusion}
This survey provides a novel summary of the existing works on blockchain scalability from the architecture, data, and protocol perspectives. To analyze the techniques for improving blockchain scalability more clearly, we classified the existing works innovatively into inner-chain and inter-chain categories within each section. Finally, we summarized the existing efforts in evaluating scalability to validate the effectiveness of the scalability improvements. We hope that this survey can help readers gain a comprehensive understanding of blockchain scalability, encourage further exploration of strategies to enhance blockchain scalability, and contribute to the development of blockchain technology.

\bibliographystyle{elsarticle-harv} 
\bibliography{reference}

\end{document}